\documentclass[prx,aps,twocolumn,superscriptaddress]{revtex4}

\usepackage{dcolumn}
\usepackage{bm}
\usepackage{amssymb}
\usepackage{color}
\usepackage[pdftex]{graphicx}

\usepackage{amsmath}
\usepackage[linesnumbered,ruled]{algorithm2e}
\usepackage{mathrsfs}
\usepackage[mathscr]{euscript}

\usepackage{url}
\usepackage[section]{placeins}
\usepackage[colorlinks=true,linkcolor=blue,citecolor=blue]{hyperref}

\begin{document}



\title{Atomistic simulation of Mott transition in fluid metals: \\ Combining molecular dynamics with dynamical mean-field theory}


\author{Zhijie Fan}
\affiliation{Department of Physics, University of Virginia, Charlottesville, VA 22904, USA}

\author{Gia-Wei Chern}
\affiliation{Department of Physics, University of Virginia, Charlottesville, VA 22904, USA}

\date{\today}

\begin{abstract}
We present a new quantum molecular dynamics (MD) method where the electronic structure and atomic forces are solved by a real-space dynamical mean-field theory (DMFT). Contrary to most quantum MD methods that are based on effective single-particle wave functions, the DMFT approach is able to describe correlation-induced Mott metal-insulator transitions and the associated incoherent  electronic excitations in an atomic liquid. We apply the DMFT-MD method to study Mott transitions in an atomic liquid model which can be viewed as the liquid-state generalization of the Hubbard model. The half-filled Hubbard liquids also provide a minimum model for alkali fluid metals. Our simulations uncover two distinct types of Mott transition depending on the atomic bonding and short-range structures in the electronically delocalized phase. In the first scenario where atoms tend to form dimers, increasing the Hubbard repulsion gives rise to a transition from a molecular insulator to an atomic insulator with a small window of enhanced metallicity in the vicinity of the localization transition. On the other hand, for Hubbard liquids with atoms forming large conducting clusters, the localization of electrons leads to the fragmentation of clusters and is intimately related to the liquid-gas transition of atoms. Implications of our results for metal-insulator transitions in fluid alkali metals are discussed.  
\end{abstract}



\maketitle

\section{Introduction}

\label{sec:intro}

Metal-insulator transition (MIT) continues to be an important subject in modern condensed matter physics even after more than sixty years of study~\cite{mott90,imada98,dobrosavljevic12}. Unlike conventional phase transitions that can be characterized by broken symmetries, MIT originates from the different dynamical behaviors of electrons, namely, itinerancy versus localization. There are several mechanisms that could cause a transition to an insulating state. For example, a band-insulator could result from a gapped Fermi surface in the presence of a long-range order through the Slater mechanism. In the absence of conventional symmetry breaking, intrinsic MIT that is caused by the localization of electrons could result from strong disorder or electron correlation. The first scenario, also called the Anderson localization, is due to a quantum-mechanical interference effect that results in well localized wave functions even in the absence of interactions~\cite{anderson58,lee85}. On the other hand, in the Mott transition scenario, localization of electrons is driven by strong short-range Coulomb repulsion~\cite{peierls37,mott49}. The interplay of these two mechanisms lead to interesting phenomena such as disorder-induced local moment formation and electronic Griffiths phase~\cite{paalanen88,milovanovic89,bhatt92,belitz94,dobrosavljevic97,miranda01,miranda05,yamamoto20}.

While extensive efforts have been devoted to understanding the effects of quenched disorder on correlated lattice models, much less is known about Mott transitions in atomic liquids, which in a sense can be viewed as correlated electrons subject to a dynamical disorder. In fact, fluid systems, such as liquid mercury and alkali metals, had played a crucial role in our understanding of MIT in disordered medium~\cite{kikoin66,mott66,hensel68,yonezawa82,hensel89}.  Early theoretical models, however, assumed that electron interactions are negligible in atomic liquids and focused on the disorder effect~\cite{mott66,cohen74,ziman61}. Assuming no particular short-range order of atoms, such approach to MIT in liquid is not much different from those developed for the MIT in amorphous solids.  For example, Mott's theory for liquid semiconductors and mercury posited a disorder-induced pseudogap within which the electron states are localized through the Anderson mechanism~\cite{mott66}. A metal-to-insulator transition occurs when the density of states at the Fermi level is below a threshold and the pseudogap opens near the Fermi energy~\cite{mott90}. 

Theoretical approaches to liquid-state MIT based on Anderson localization mechanism often implicitly assumes an atomic configuration similar to that of amorphous solid, an assumption which is not always justified. On the other hand, models based on percolation theory have been proposed that link the atomic structure to the MIT in liquid. For example, lattice gas system has been used to model the percolation MIT in the super-critical regime of fluid metals~\cite{kirkpatrick73,phelps76,odagaki75,franz84}. Percolation scenarios were also proposed based on the reverse Monte Carlo modeling of neutron diffraction data of expanded alkali fluid~\cite{nield91}. As the system approaches the liquid-gas critical point with a decreasing atomic density, a conducting network persists where finite clusters are connected through weak atomic links. Since these finite-sized clusters are stabilized by the delocalization of electrons through metallic bonding, the resultant MIT is suggested to be driven by a process similar to bond percolation~\cite{nield91,tarazona95,arai99}.


The importance of disorder for electron localization in liquid metals does not preclude the correlation effect also as a driving force for MIT,   especially for transition-metal or rare-earth compounds. For example, the first-order phase transition in liquid metallic cerium is mainly driven by the localization of $f$ electrons~\cite{cadien13}. Moreover, in the percolation picture discussed above, finite atomic clusters stabilized by the binding force from itinerant electrons could be rather resilient during a density-driven MIT in an atomic liquid. Electron localization driven by correlation effects thus could be the dominant mechanism for the break up of atomic clusters in this scenario. Of particular interest is the MIT in expanded metallic alkali fluids. Early theoretical works already emphasized the importance of electron correlation for the MIT in such monovalent liquid systems~\cite{yonezawa73,yonezawa74}. Various experimental studies also hinted at a correlation-driven metal-nonmetal transition in supercritical alkali liquid~\cite{jungst85,winter87,hensel85,hensel89b,freyland79,hanany83,warren89}. In particular, experiments on liquid cesium and rubidium observed an enhancement of magnetic susceptibility close to the critical density~\cite{freyland79,hanany83}, indicating the emergence of localized magnetic moments which is a telltale sign of Mott-Hubbard type transition.

Perhaps the most unique aspect of MIT in a liquid state is the interplay between atomic dynamics and electron localization. The electronic structure and transport properties naturally depend on the short-range atomic configuration. On the other hand, the structure of the atomic liquid is determined by the interatomic forces, which in turn depend on the electronic properties and particularly the degree of electron delocalization. A case in point is the persistence of finite metallic clusters embedded in a percolating network of an expanded fluid, as discussed above. Another example is the dissociation of molecules due to a correlation-driven localization of electrons. In order to account for the subtle interrelationship between atomic configuration and electronic structure, computational methods such as Monte Carlo or molecular dynamics (MD) are required to account for the dynamical nature of the atoms.


In this paper, we present a novel quantum molecular dynamics (QMD) method that can properly take into account the various aspects of metal to non-metal transitions in an atomic liquid discussed above. As in standard QMD methods~\cite{marx09}, the forces that drive the motion of atoms are computed from solutions of the electronic structure problem at every time-step. Notably, in order to describe the strong electron-correlation effects and the resultant Mott-Hubbard type transitions, the electronic system is solved by a real-space dynamical mean-field theory (DMFT) at every time-step. Moreover, contrary to most QMD methods which reduce the many-electron problem into an effective single-particle Schr\"odinger equation, the DMFT approach aims to solve self-consistent electron Green's functions, and hence is able to provide incoherent electronic excitations that are characteristic of Mott transitions.



The DMFT-MD method is used to simulate the MIT in a model fluid metal with a correlated $s$-band. Formulated within the framework of tight-binding molecular dynamics (TBMD), this liquid model offers a natural and intuitive generalization of the Hubbard model to an atomic liquid, allowing one to study the generic interplay of dynamical disorder and electron correlation. It is worth noting that this Hubbard liquid model at half-filling also serves as a minimum model for MIT in alkali fluids. Depending on the atomic bonding structures in the itinerant phase, the Hubbard liquid model exhibits two distinct types of correlation-driven transition to the Mott phase depending on the range of the core repulsion relative to that of the electron hopping. 

For Hubbard liquids with a longer-ranged repulsive core, delocalized electrons give rise to the formation of diatomic molecules similar to the molecular hydrogen gas at high temperatures. It is worth noting that this molecular fluid is in an insulating phase as the electrons are confined to individual pairs of atoms. Upon increasing the on-site Coulomb repulsion $U$, the correlation-induced localization of electrons leads to the break up of the covalent bond and the dissociation of the molecule. The Mott insulating state at large $U$ thus is comprised of isolated atoms with predominantly a short-range repulsive interatomic interaction. Interestingly, a small window of enhanced metallicity emerges in between the molecular and atomic gas phases both of which are electrically insulating. 

In the second scenario which corresponds to atoms with a short-ranged repulsive core, the electron-mediated cohesive force results in extensive clusters of atoms. As the atomic clusters are stabilized by a metallic-like bonding, the resultant liquid phase is metallic. The Mott transition in this system thus offers a generic representative example of correlation-driven metal-to-insulator transitions in an atomic liquid. In contrast to Mott transitions in both crystalline and amorphous systems, the interplay between the mobile atoms and electron localization gives rise to interesting features that are unique to MIT in liquid state. In particular, the persistence of metallic clusters highlights the importance of atomic dynamics in the MIT of an expanding fluid.

The rest of the paper is organized as follows. In Sec.~\ref{sec:model} we present a model system that generalizes the well-studied Hubbard model to an atomic liquid. This so-called Hubbard liquid model is formulated as a TBMD model with the addition of an on-site Hubbard interaction. The {\em ab initio} derivation of the model parameters is outlined and the relevance to alkali liquids is discussed. The method of DMFT-based MD simulation is then discussed in Section~\ref{sec:dmft-md}. We first discuss the formulation of quantum MD methods that are based entirely on the electron Green's function. This is followed by a review of the real-space DMFT method for inhomogeneous correlated electron systems. The results of the DMFT-MD simulation on the Hubbard liquid model are presented in Sec.~\ref{sec:results}. In particular, we present the first-ever numerical simulation of the evolution of electron spectral function and optical conductivity during a liquid-state MIT. Finally, Section~\ref{sec:discussion} summarizes our work.

\section{The Hubbard liquid model}

\label{sec:model}

The repulsive-interaction Hubbard model~\cite{hubbard63,hubbard64,kanamori63,gutzwiller63} is one of the canonical models for strongly correlated electron systems.  Despite its seemingly simplicity, the single-band Hubbard model exhibits a wide range of correlated electron behavior including interaction-driven metal-insulator transitions, superconductivity, and magnetism~\cite{tasaki98,fazekas99}.
Originally proposed as a model system to describe correlation-driven MIT, a renewed interest in the 2D Hubbard model was spurred by the discovery of high-temperature cuprate superconductors~\cite{arovas21}.
A wide variety of state-of-the-art numerical methods have also been applied to study the phase diagram of the 2D square-lattice Hubbard model~\cite{leblanc15,qin20,schafer21}.
The Hubbard model can be solved exactly only in the one-dimensional case, although no Mott transition is found in this special limit~\cite{lieb68}. On the other hand, Mott transition has been demonstrated in the infinite dimensional limit of the Hubbard model~\cite{metzner87,metzner89,hartmann89}, which also inspires the development of the approximation methods such as the Gutzwiller approximation (GA) and DMFT.  In particular, the frustrated Hubbard model at half-filling is the testbed for studying the correlation-induced MIT. 

Here we present a minimum generalization of the Hubbard model to an atomic liquid, which is formulated within the tight-binding molecular dynamics (TBMD) framework. The Hamiltonian governing the atomic dynamics in TBMD consists of three parts
\begin{eqnarray}
	\label{eq:TBMD}
	\mathcal{H} = \sum_i \frac{\mathbf P_i^2}{2 M} + E_{\rm bind}(\{\mathbf R_i \}) 
	+  \frac{1}{2} \sum_{i \neq j} \phi( \left| \mathbf R_i - \mathbf R_j \right|),
\end{eqnarray}
where $\mathbf R_i$ and $\mathbf P_i$ are the position and momentum vectors, respectively, of the $i$-th atom. The first term above describes the classical kinetic energy of atoms with mass~$M$. The second term represents the electronic binding energy, or ``band" energy. In the TBMD formalism, it is basically given by the total energy of occupied eigenstates of a tight-binding (TB) Hamiltonian $\mathcal{H}_{\rm TB}$, i.e. $E_{\rm bind} = \sum_m f_m \langle \psi_m | {\mathcal{H}}_{\rm TB} (\{\mathbf R_i \}) | \psi_m \rangle$, where  $|\psi_m\rangle$ is the $m$-th eigenstate and $f_m$ is the corresponding Fermi-Dirac occupation factor. Given a set of localized orbitals, the TB Hamiltonian is then parametrized by the hopping integrals $h_{ij}$. The $\phi_{ij} = \phi( \left| \mathbf R_i - \mathbf R_j \right|)$ in the third term is a pairwise potential representing the ion-ion repulsion and the correction for double counting the electron-electron interaction in the second term [...]. The functional forms of the hopping integrals as well as the pair potential are determined via fitting with either experimental data or first-principles calculations. 

To include the electron correlation effect more explicitly, here the electronic energy $E_{\rm elec}$ is assumed to be computed from the expectation value of a Hubbard Hamiltonian. The total Hamiltonian for both atoms and electrons is then given by 
\begin{eqnarray}
	\label{eq:H}
	\mathcal{H} &=& \sum_{ij} \sum_{\sigma} h( \left| \mathbf R_i - \mathbf R_j \right|) \,  c^\dagger_{i \sigma} c^{\,}_{j \sigma}  + U \sum_i n_{i  \uparrow} n_{i  \downarrow}  \nonumber \\
	& & + \sum_i \frac{\mathbf P_i^2}{2 M}  + \frac{1}{2} \sum_{i \neq j} \phi( \left| \mathbf R_i - \mathbf R_j \right|).
\end{eqnarray}
Here $c^\dagger_{i, \sigma}$ ($c^{\,}_{i,\sigma}$) creates (annihilates) an electron of spin $\sigma = \uparrow, \downarrow$ at atom-$i$, $n_{i,\sigma} = c^\dagger_{i,\sigma} c^{\,}_{i, \sigma}$ is the electron number operator, $h_{ij} = h(|\mathbf R_i - \mathbf R_j|)$ denotes the hopping amplitude between a pair $(ij)$ of atoms, $U > 0$ is the Hubbard term due to on-site Coulomb repulsion. 
The first two terms in Eq.~(\ref{eq:H}) correspond to a Hubbard model with random hoppings $h_{ij}$ that are determined by the instantaneous atomic configuration $\{\mathbf R_i\}$.  For clarity, here we use the uppercase letters to denote position, momentum, and mass for nuclei, while lowercase letters are used for electrons.


The parameters of the Hubbard liquid model can be derived from first-principles calculations. For example, in the {\em ab initio} tight-binding and TBMD methods, predetermined functions for the hopping integral and pair potential are determined through fitting to density function theory (DFT) calculations~\cite{andersen84,sutton88,wang08,elstner98,horsfield00,khan89,wang89,goedecker94,goodwin89}. Modern {\em ab initio} approach to build TB Hamiltonians and Hubbard-type models, as often employed in the LDA+$U$ or LDA+DMFT methods, is based on the Wannier functions~\cite{wannier37,ashcroft76,mazari97,marzari12}. These localized basis functions for given correlated bands are obtained by downfolding the KS Hamiltonians often solved  using the plane-wave basis. Attempts have also been made to the construction of Wannier functions in disordered systems~\cite{kohn73,kivelson82,zhou10,silvestrelli98}.  

The determination of the Hubbard $U$ parameter is a challenging task for such DFT based methods. Yet another first-principles approach is to use well-defined atomic-like orbitals to build the Wannier functions~\cite{vidal91,ortega98,spalek92,spalek07}. This method has the advantage that the Hubbard parameter can be unambiguously determined. In practical calculations, Gaussian-type orbitals are often used to approximate the atomic basis functions in order to efficiently carry out the multiple integrals listed above, as well as the force calculations required for MD simulations~\cite{hehre69}. However, even with the Gaussian functions, the calculation of parameters for the instantaneous TB-Hubbard model is still computationally very expensive.


Let $w_i(\mathbf r)$ is the Wannier function centered at the $i$-th atom, the electron hopping in Eq.~(\ref{eq:H}) is given by the integral 
\begin{eqnarray}
	\label{eq:h-integral}
	h_{ij} = \int w^*_i(\mathbf r) \left[ -\frac{\hbar^2 \bm\nabla^2}{2 m} + \sum_k V_k(\mathbf r) \right] w_j(\mathbf r) \, d^3\mathbf r
\end{eqnarray}
where $m$ is the electron mass, and $V_{k}(\mathbf r)$ is the pseudo-potential of the $k$-th ion that accounts for the Coulomb attraction $-Z e^2/|\mathbf r - \mathbf R_k|$ screened by the core electrons. Generally speaking, the Wannier functions are well localized at the individual atoms, which means the transfer integral $h_{ij}$ decays rapidly with increasing separation~$R_{ij}$. 



The diagonal term of the integral in Eq.~(\ref{eq:h-integral}) can be expressed as $h_{ii} = \epsilon_i + \sum_{k\neq i} \mathcal{E}_{i, k}$, where the first term
\begin{eqnarray}
	\epsilon_i =  \int w^*_i(\mathbf r) \left[ -\frac{\hbar^2 \bm\nabla^2}{2 m} + V_i(\mathbf r) \right] w_i(\mathbf r) \, d^3\mathbf r , 
\end{eqnarray}
contributes a diagonal on-site energies $ \sum_{i,\sigma} \epsilon_i c^\dagger_{i\sigma} c^{\,}_{i \sigma} $ to the Hamiltonian in Eq.~(\ref{eq:H}). The variation of this on-site energy is due to the non-uniformity of the Wannier functions. However, the non-uniformity is expected to be small because of the charge-neutrality condition for individual atoms. And since a constant $\epsilon$ just shifts the energy reference, this diagonal term is neglected in the Hamiltonian for convenience.   The second part of $h_{ii}$ is given by the integral
\begin{eqnarray}
	\mathcal{E}_{i, k} & = & \int V_k(\mathbf r) \left| w_i(\mathbf r) \right|^2  \, d^3\mathbf r,
\end{eqnarray}
which is the interaction energy, through a screened Coulomb potential, between the $k$-th ion and the electron localized at atom-$i$. 

The Coulomb repulsion between electrons, in the second quantization formulation, is given by four-fermion interactions $\sum_{ijkl} \mathcal{V}_{ijkl} \, c^\dagger_{i\sigma} c^\dagger_{j \sigma'} c^{\,}_{k \sigma'} c^{\,}_{l \sigma}$, where the interaction parameters are
\begin{eqnarray}
	\mathcal{V}_{ij; kl} = \int \int \frac{e^2 w^*_i(\mathbf r) w^*_j(\mathbf r') w_k(\mathbf r) w_l(\mathbf r')}{|\mathbf r - \mathbf r'|} \, d^3\mathbf r \,d^3\mathbf r'.
\end{eqnarray}
The Hubbard liquid Hamiltonian Eq.~(\ref{eq:H}) only includes the on-site repulsion explicitly; the corresponding Hubbard parameter, which in general depends on atom index, is given by $U_i = \mathcal{V}_{ii; ii}$. 
The off-site Coulomb interactions such as~$\mathcal{V}_{ij; ij}$ can be accounted for through a conventional Hartree-Fock treatment. Specifically, they are included in the effective single-electron Schr\"odinger equation, from which the Wannier functions $w_i(\mathbf r)$ are obtained. The double-counting terms of the HF calculation is included in the classical pair potential discussed below. 

With the Hubbard $U$ explicitly separated from other electron-electron interactions, the pair potential is expressed as
\begin{eqnarray}
	\phi_{ij} = \frac{e^2}{|\mathbf R_i - \mathbf R_j|} - \left(\mathcal{E}_{i,j} + \mathcal{E}_{j, i} \right) + \mathcal{V}_{ij; ij} - \frac{1}{2}\mathcal{V}_{ij; ji}.
\end{eqnarray}
The dominant contribution is the classical Coulomb repulsion between the positively charged nuclei. The two $\mathcal{E}$-terms describe the ion-electron interaction, the Hartree term $V_{ij; ij}$ accounts for the electron-electron Coulomb repulsion, and the last term is the mean-field exchange interaction. 
At large separation, the first three terms correspond to an approximate Coulomb interaction: 
\begin{eqnarray}
	\phi_{ij} \approx \int \int \frac{\rho_i(\mathbf r) \rho_j(\mathbf r') }{ |\mathbf r - \mathbf r'| }d^3\mathbf r d^3\mathbf r' + \cdots, 
\end{eqnarray}
where $\rho_i(\mathbf r) = \delta(\mathbf r - \mathbf R_i) - |w_i(\mathbf r)|^2$ is charge density of the nucleus at the $i$-th atom screened by the $s$-orbital electron. Here we have explicitly shown that a dominant contribution to the pair potential is the Coulomb interaction of the screened nuclei charge $\rho_i(\mathbf r)$. Since the net charge associated with $\rho_i(\mathbf r)$ is zero, the integral $\phi_{ij}$ also decays rapidly with increasing atomic distance $R_{ij}$.

Although these model parameters can be obtained from first principles as outlined above, such calculations, which have to be performed at every MD time-step, is computationally very expensive. 
Instead, we treat the Hubbard liquid model Eq.~(\ref{eq:H}) as a minimum model system to capture the Mott-Hubbard physics in an atomic liquid system and investigate the new physics that arise from the interplay between electron correlation and atomic dynamics. To this end, further approximations are introduced. First, the Hubbard parameter $U$ is treated as a model parameter, which is independent of atoms. Explicit calculations in the case of hydrogen liquid show that the variation of $U$ among atoms is indeed small~\cite{chen-x}. 

Second, as discussed above, both the hopping integral and the pair potential are strongly decaying functions of distance  thanks to the well localization of Wannier functions. Although many functional forms have been proposed~\cite{goodwin89,varshni57}, for simplicity we assume that the hopping integral decays exponentially with the distance $h(R) = h_0 \exp(-R/\xi)$, where $h_0$ and $\xi$ are two model parameters characterizing the amplitude and range of electron hopping. We note in passing that similar exponential function is widely used for modeling MIT in doped semiconductors~\cite{milovanovic89,bhatt92}. 
On the other hand, in order to describe repulsive interaction of different ranges, we consider the following functional form for the pair potential $\phi(R) = \phi_0 \,\exp[ -  (R / \ell) - b( R/\ell)^4  ]$. Here $\phi_0$, $\ell$, and $b$ are also parameters of the model system, with a larger parameter $b$ corresponding to a sharper repulsive core. 






It is worth noting that, because of these approximations, our results are not expected to quantitatively describe MIT in realistic alkali fluids or other liquid metals. Instead, similar to the spirit of the lattice Hubbard model, our goal is to study the behaviors, both atomically and electronically  of this new model correlated electron system under different parameters. In particular, this work is aimed to provide a fundamental picture of Mott metal-insulator transition in an atomic liquid. Moreover, we note that the Hubbard liquid model in Eq.~(\ref{eq:H}) not only is conceptually a natural generalization of the well-studied lattice version, but also serves as a minimum model, especially at half-filling, for the metal-insulator transition in alkali fluids. 


\section{molecular dynamics with Dynamical mean-field theory}
\label{sec:dmft-md}

Quantum MD or more specifically {\em ab initio} MD can be classified into adiabatic or non-adiabatic approaches~\cite{marx09}. The representative example of adiabatic QMD is the Born-Oppenheimer molecular dynamics (BOMD) which assumes that electrons adjust instantaneously to the slower motion of the nuclei so that the motion of the latter is governed by a single adiabatic potential energy surface (PES). On the other hand, electronic transitions between multiple PESs are taken into account in non-adiabatic MD simulations~\cite{tully12}. In this work, we will focus on the adiabatic limit, which is also the starting point of most quantum MD methods.

\subsection{Quantum molecular dynamics: A brief overview} 

The evolution of atomic configuration can be efficiently modeled by the molecular dynamics (MD) simulation~\cite{binder04,allen89,rapaport04}, which has a long history dating back to the famous Fermi-Pasta-Ulam-Tsingou nonlinear chain simulation in 1953~\cite{fermi55}. By providing a general approach to understand and analyze material functionalities in terms of dynamics at the atomic level, MD essentially serves the role of a computational microscope. Although conceptually MD simulation is simply the integration of Newton equation of motion for a large number of atoms, the challenging part is the calculation of interatomic forces. In the widely used classical MD methods, these interatomic forces are computed from predetermined empirical potentials or force fields~\cite{mackerell04}. Since the evolution of the electronic subsystem is not consistently accounted for, such classical approaches certainly cannot describe electronic phase transitions such as MIT.

The scope and predictive power of MD methods are greatly enhanced by using the quantum approach for force calculation~\cite{car85,payne92,kresse93,tuckerman02,marx09,attaccalite08}. In such quantum MD (QMD) schemes, the forces acting on atoms are obtained by solving the many-electron Schr\"odinger equation on the fly as the atomic trajectories are generated. The validity and limitation of a QMD scheme then depend on the approximations used in solving the many-body problem. For example, the well known Hartree-Fock (HF) mean-field methods were used in the early development of {\em ab~initio} MD methods~\cite{gordon88,field91}. The most popular QMD methods  nowadays are based on density functional theory (DFT)~\cite{hohenberg64,kohn65,jones15}, or more specifically the Kohn-Sham (KS) approach. By recasting the intractable complexity of the many-electron interactions into the form of an effective one-electron energy that is a unique functional of the electron density, the KS method achieves a desirable tradeoff between accuracy and efficiency. Assisted by high-performance computers and advanced algorithms, DFT-MD is now firmly established as an essential research tool in physics, chemistry, biology, and materials sciences.

Although DFT is in principle exact, its accuracy in practical implementation depends on approximations used for the exchange-correlation functional, whose exact universal expression is not known. The local density approximation (LDA) and its variants are among the most popular methods~\cite{perdew91,becke88,perdew96}. One particular limitation of these approximations is its inability to describe phenomena that is due to strong electron correlation. While modified methods such the self-interaction correction or DFT+$U$ have proven useful for some applications~\cite{perdew81,tsuneda14,anisimov91,anisimov91b}, it is still very difficult to capture generic electron correlation effects and particularly the Mott metal-insulator transition in the KS approach. 

On the other hand, several many-body techniques have been developed to handle strong electron correlation in lattice models such as the Hubbard and Anderson Hamiltonians~\cite{fazekas99,hewson93}. Among them, the Gutzwiller variational method is perhaps the most efficient approach that successfully capture the essential correlation effects~\cite{gutzwiller63,gutzwiller64,gutzwiller65}. For example, the Brinkman-Rice theory~\cite{brinkman70} of the Hubbard model, which is based on the Gutzwiller method, had offered important insight on the Mott transition. 
Contrary to a single Slater-determinant underlying either the HF or KS methods, the Gutzwiller wave function is a multi-Slater-determinant that is variationally optimized to balance the kinetic energy gain due to electron delocalization against the local Coulomb repulsion when two electrons reside at the same orbital~\cite{vollhardt84}.

In an effort to develop MD methods for correlated electron materials, a new QMD scheme~\cite{chern17,suwa19,chern19} was recently proposed that is based on the Gutzwiller wave function and the Gutzwiller approximation (GA). Interatomic forces in such GA-MD simulations are computed from a Gutzwiller wave function that has to be iteratively optimized at every time-step. Crucially, similar to other self-consistent approaches such as HF or KS methods, the GA also reduces the intractable many-body problem into an effective single-electron one in terms of a renormalized tight-binding Hamiltonian, which is to be solved self-consistently. Indeed, in its modern formulation in terms of slave bosons, GA can be viewed as a mean-field theory for the Mott transition with amplitudes of slave-bosons serving as the order parameters~\cite{kotliar86,bunemann07,lechermann07,lanata17}. Consequently, the computational cost of GA-MD simulation is similar to that of DFT-MD.

It is worth noting that MD methods based on self-consistent independent-electron approach, including HF, KS, and GA, can only describe the behavior of quasi-particles of the electronic system. The quasi-particles basically are solutions of the self-consistent single-electron Schr\"odinger equation. Importantly, while the HF or KS quasi-particles are electrons whose single-particle states have been renormalized by interactions, but whose effective mass and Fermi distribution remain unchanged with respect to the non-interacting case, the quasi-particles described by the GA are Landau quasi-particles with an enhanced mass $m^*$~\cite{vollhardt84};  the divergence of $m^*$ signals the localization of electrons and the onset of Mott transition.

The quasi-particle weight in GA is directly related to the renormalization of inter-site hopping and the electron bandwidth. Although GA provides a qualitatively, and often quantitatively, correct description for correlated metals through the concept of renormalized quasi-particles, it fails to account for the incoherent electronic excitations and the appearance of Hubbard bands. Moreover, the non-Fermi liquid behavior and electronic Griffiths phase~\cite{miranda01,miranda05,yamamoto20} that result from the interplay of disorder and electron correlation are also beyond the capability of GA-MD methods.

In this work, we propose a new quantum MD scheme that goes beyond self-consistent independent-electron picture for strongly correlated electron systems. Our approach is based on an efficient integration of tight-binding molecular dynamics method with the dynamical mean field theory (DMFT)~\cite{kotliar04,georges96,kotliar06}. Contrary to most quantum MD methods that rely on solving an effective single-particle Schr\"odinger equation, the DMFT-MD scheme is entirely based on the self-consistent solution for the electron Green's function, thus offering the capability of computing the complete spectral function that includes the quasi-particle peaks and the incoherent excitations represented by the Hubbard bands. Within DMFT, the Mott transition results from the transfer of spectral weight from the quasiparticle peak to the Hubbard bands.

\subsection{Quantum molecular dynamics based on electron Green's functions}

\label{sec:adiabatic}

The governing equations of adiabatic MD are often obtained from certain ansatz for the total wave function, as detailed in Ref.~\cite{marx09}. Such approaches have also been the basis of most QMD simulations that reply on the effective independent-electron methods. Since the DMFT-MD is formulated in terms of electron Green's functions, instead of wave functions, we first discuss such QMD formalism and outline the derivation of the corresponding adiabatic approximation.  We start with the dynamical equation for the nuclei. In order to properly obtain the electronic force, we consider the Heisenberg equation of motion for the nuclei position operator $d\hat{\mathbf R}/dt = [\hat{\mathbf R}, \mathcal{H}] / i\hbar$. Following standard procedures, the expectation value of this equation gives the classical Newton equation of motion
\begin{eqnarray}
	\label{eq:newton}
	M \frac{d^2 \mathbf R_i}{dt^2} + \gamma \frac{d \mathbf R_i}{dt} = -\left\langle \frac{\partial \mathcal{H}}{\partial \mathbf R_i} \right\rangle +  \bm\eta_i(t),
\end{eqnarray}
where $M$ is mass of the nuclei, $\mathbf R_i = \langle \hat{\mathbf R}_i \rangle$ is now a classical position vector. We have included the dissipation force and thermal noise as in the Langevin dynamics; here $\gamma$ is a damping coefficient and $\bm\eta_i(t) = (\eta_i^x, \eta_i^y, \eta_i^z)$ denotes a vector whose components are normal-distributed random variables with zero mean:
\begin{eqnarray}
	\langle \eta^\alpha_i(t) \rangle &=& 0, \\
	\langle \eta^\alpha_i(t) \eta^\beta_j(t) \rangle &=& 2\gamma k_B T \delta_{\alpha\beta} \delta_{ij} \delta(t - t'). \nonumber
\end{eqnarray}
The standard second-order velocity-Verlet method is used to integrate the above equation of motion. It is worth noting that the deterministic force, the first term on the right hand side of Eq.~(\ref{eq:newton}), is not given by the conservative form $-\partial \langle \mathcal{H} \rangle / \partial \mathbf R_i$. The equivalence of these two expressions is due to Hellmann-Feynman theorem under certain approximations~\cite{hellmann37,feynman39}. 
Substituting the Hamiltonian Eq.~(\ref{eq:H}) into the above expression, we obtain two contributions to the force
\begin{eqnarray}
	\mathbf F_i = - \left\langle \frac{\partial \mathcal{H}}{\partial \mathbf R_i} \right\rangle = \mathbf F_i^p + \mathbf F_i^e,
\end{eqnarray}
where the superscript $p$ and $e$ indicates contributions from the pair potential and the electrons, respectively. The first term, which describes the short-range repulsion between atoms, is given by
\begin{eqnarray}
	\mathbf F_i^p = -\sum_j \frac{\partial \phi_{ij}}{\partial \mathbf R_i} = \sum_j \phi'(R_{ij}) \,\hat{\mathbf n}_{ij},
\end{eqnarray}
where $\phi'(R) = d\phi(R)/dR$ and $\hat{\mathbf n}_{ij}$ is a unit vector pointing in the direction of vector $\mathbf R_{ij} = \mathbf R_j - \mathbf R_i$. We note that in deriving this result, we have implicitly assumed $\langle f(\hat{\mathbf R}_i) \rangle = f\bigl(\langle \hat{\mathbf R}_i \rangle) = f(\mathbf R_i)$, where $f(x)$ is an arbitrary function. This, of course, is the classical approximation for nucleic degrees of freedom, which is the fundamental assumption of MD methods.  This classical force is easy to compute for the Hubbard liquid model. The electron part is 
\begin{eqnarray}
	\mathbf F^e_i &=& -\sum_{j, \sigma} \frac{\partial h_{ij}}{\partial \mathbf R_i} \langle (c^\dagger_{i \sigma} c^{\,}_{j \sigma} + {\rm h.c.} ) \rangle \nonumber \\
	&=& 2 \sum_{j, \sigma} h'(R_{ij}) \, \hat{\mathbf n}_{ij} \, {\rm Re}[ \rho_{j \sigma, i \sigma} ].
\end{eqnarray}
where  $h'(R) = dh(R)/dR$.
The electronic force depends on the single-electron (reduced) density matrix
\begin{eqnarray}
	\rho_{i\sigma, j\sigma'}(t) = \langle c^\dagger_{j \sigma'}(t) c^{\,}_{i \sigma}(t) \rangle.
\end{eqnarray}
It is worth noting that both forces as well as the reduced density matrix vary with time. For effective independent-electron approaches, such as HF or GA, this density matrix is computed from the eigen-solution of the effective one-particle Hamitlonian $H^{\rm eff}_{i \sigma, j \sigma'}[\{\mathbf R_i(t) \}]$, which depends on time through the nuclei coordinates. Specifically, let $U^{m}_{i, \sigma}$ be the normalized eigenvector of the one-electron Hamiltonian with eigenenergy $\epsilon_m$, the density matrix is $\rho_{i \sigma, j \sigma'} = \sum_{m} f(\epsilon_m) U^{m\,*}_{i \sigma} U^{m}_{j \sigma'}$, where $f(\epsilon_m)$ is the Fermi-Dirac function for the occupation probability of the $m$-th eigenstate.

As discussed above, contrary to the effective independent-electron QMD methods, the DMFT-MD is formulated in terms of the electron Green's function. Of particular importance is the lesser Green's function 
\begin{eqnarray}
	{G}^{<}_{i\sigma, j \sigma'}(t_1, t_2) = i \langle c^\dagger_{j \sigma'}(t_2) c^{\,}_{i \sigma}(t_1) \rangle.
\end{eqnarray}
The reduced density matrix is given by the equal-time lesser Green's function: $\rho_{i \sigma, j \sigma'}(t) = -i {G}^<_{i \sigma, j \sigma'}(t, t)$. Also important are the retarded Green's function defined
\begin{eqnarray}
	{G}^R_{i \sigma, j \sigma'}(t_1, t_2) = -i \theta(t_1-t_2) \langle \{ c^{\,}_{i \sigma}(t_1), c^\dagger_{j \sigma'}(t_2) \} \rangle, 
\end{eqnarray}
and the associated advanced Green's function ${G}^A_{i\sigma,j \sigma'}(t_1, t_2) = [{G}^R_{j \sigma', i \sigma}(t_2, t_1) ]^\dagger$.
In terms of the lesser Green's function, the electronic force becomes
\begin{eqnarray}
	\label{eq:force1}
	\mathbf F^e_i(t) = 2 \sum_{j, \sigma} h'(R_{ij}(t)) \, \hat{\mathbf n}_{ij}(t) \, {\rm Im}\,{G}^<_{j\sigma, i \sigma}(t, t).
\end{eqnarray}
Here we have explicitly included all the time dependences. A consistent dynamical description thus requires the equation of motion for the lesser Green's function. This can be achieved by the nonequilibrium Green's function (NEGF) theory. For example, a complete dynamical theory can be obtained by combining the Newton equation of motion~(\ref{eq:newton}) for nuclei with the Kadanoff-Baym equations for the Green's functions. 
\begin{eqnarray}
	\label{eq:KB1}
	& &  \left[ i \frac{\partial}{\partial t_1} + \mu   - {\mathbf h}(t_1) \right] \mathbf G^<(t_1, t_2)  =  \\
	& & \qquad \int dt_3 \Bigl[ \boldsymbol{\Sigma}^R(t_1, t_3) \mathbf G^<(t_3, t_2) 
	  + \boldsymbol{\Sigma}^<(t_1, t_3) \mathbf G^A(t_3, t_2) \Bigr], \nonumber
\end{eqnarray}
\begin{eqnarray}
	\label{eq:KB2}
	& &\left[ i \frac{\partial }{\partial t_1} +\mu - \mathbf h(t_1) \right] \mathbf G^{R/A}(t_1, t_2) 
	= \delta(t_1 - t_2) \, {\mathbf I}  \qquad \quad \\
	& & \qquad \qquad \qquad + \int dt_3 \boldsymbol{\Sigma}^{R/A}(t_1, t_3) \mathbf G^{R/A}(t_3, t_2).  \nonumber
\end{eqnarray}
Here bold symbols are used to denote matrices in the atom-spin basis, $\boldsymbol{\Sigma}^<$ and $\boldsymbol{\Sigma}^{R/A}$ are the lesser, retarded, and advanced self-energies, respectively, $\mu$ is the electron chemical potential, and the time-varying matrix~$\mathbf h(t)$ describes the electron hopping of the Hamiltonian in~Eq.~(\ref{eq:H}), 
\begin{eqnarray}
	 h_{i \sigma, j \sigma'}(t) = \delta_{\sigma\sigma'} \, h(|\mathbf R_i(t) - \mathbf R_j(t)| ).   
\end{eqnarray}	
For many-body interacting systems, after introducing suitable approximations such as Born approximation or HF to relate the self-energy $\Sigma$ to the Green's functions, the Kadanoff-Baym equation can be integrated along with the Newton equation of motion.  However, even without the atom dynamics, numerical integration of the inhomogeneous Kadanoff-Baym equation is computationally very demanding. So far its implementation is restricted to small systems~\cite{dahlen07,friesen09,schlunzen16,schlunzen20}. 

The calculation can be much simplified when the nuclei and electrons evolve on significantly different time scales, allowing the system to split into fast (electron) and slow (nuclei) degrees of freedom. To this end, we transform the Kadanoff-Baym equations to the Wigner space where fast and slow time scales are easily identifiable. We define the average and relative time parameters: $t = (t_1 + t_2)/2$ and $\tau = t_1 - t_2$, and introduce the Wigner representation of the Green's functions, 
\begin{eqnarray}
	& & {\mathbf G}(\omega, t) = \int  e^{i \omega \tau} \mathbf G\left(t_1, t_2 \right) \, d\tau, \nonumber \\
	& & \mathbf G(t_1, t_2) = \frac{1}{2\pi} \int e^{-i \omega \tau} {\mathbf G}(\omega, t)\, d\omega. 
\end{eqnarray}
A coupled set of differential equations for the Green's functions in the Wigner representation is obtained by applying the same transformation to Eqs.~(\ref{eq:KB1}) and~(\ref{eq:KB2}). Working in the Wigner space allows one to perform a systematic adiabatic expansion by using variation with respect to the central time $t$ as a small parameter. Similar approaches have recently been extensively investigated in the context of current-induced forces in quantum transport and nonequilibrium BOMD simulations~\cite{bode11,kershaw17,dou17,honeychurch19,kershaw19}.  Formally, we introduce an adiabaticity parameter $\epsilon$ which is of the order of $\epsilon \sim |d\mathbf R_i/dt|$, and expand the Green's functions in a power series, $\mathbf{G} = \mathbf{G}^{(0)} + \epsilon \mathbf{G}^{(1)} + \epsilon^2 \mathbf{G}^{(2)} + \cdots$. The Kadanoff-Baym equations can then be solved systematically for each order of $\epsilon$. 

The adiabatic limit, which is also most relevant to our work, is given by the zero-th order results. For the retarded/advanced Green's functions, the zero-th order solutions are
\begin{eqnarray}
	\label{eq:G_RA_sol0}
	{\mathbf G}^{R/A}(\omega, t) = \left[ (\omega + \mu) \,\mathbf I - \mathbf h(t)- { \bm{\Sigma}}^{R/A}(\omega, t) \right]^{-1}.  
\end{eqnarray}
Here the self-energies ${\bm{\Sigma}}^{R/A}$ only depends on the zeroth-order Green's functions through a given many-body scheme, which in our case is the DMFT to be discussed below. Notably, Eq.~(\ref{eq:G_RA_sol0}) is essentially  the ``equilibrium" Green's functions one would obtain for the  Hubbard Hamiltonian defined by the instantaneous atomic configuration $\{\mathbf R_i(t) \}$. The zeroth-order lesser Green's function, which is needed for the force calculation, is given by
\begin{eqnarray}
	{\mathbf G}^<(\omega, t) = {\mathbf G}^R(\omega, t) \, {\bm{\Sigma}}^<(\omega, t) \, {\mathbf G}^A(\omega, t),
\end{eqnarray}
which is the Keldysh equation for the lesser Green's function in the steady state. Higher-order terms in the adiabatic expansion can be found in, e.g. Refs.~\cite{bode11,kershaw19}.   

The electronic force in Eq.~(\ref{eq:force1}) can now be expressed of $\mathbf G^<$ in the Wigner representation:
\begin{eqnarray}
	\mathbf F^e_i(t) = -\frac{1}{\pi} \int d\omega \frac{\partial }{\partial \mathbf R_i} {\rm Tr}\bigl[ \mathbf h(t) \, {\rm Im}\,\mathbf G^<(\omega, t) \bigr].
\end{eqnarray}
To summarize, the adiabatic limit of the Green's functions in the Wigner representation is equivalent to imbuing a static equilibrium solution parameterized by the central time $t$. This result is consistent with the intuitive picture of Born-Oppenheimer approximation, where electrons are assumed to stay in the equilibrium state of the instantaneous Hamiltonian. Also importantly, the adiabatic expansion of the Kadanoff-Baym equations offers a systematic approach to formulate the DMFT-MD and higher-order corrections.  

\subsection{Real-space dynamical mean-field theory}

The adiabatic expansion discussed above shows that at each instance the electronic system can be treated in quasi-equilibrium. Under this condition, the various Green's functions and self-energies are related. We focus on the retarded self-energy ${\boldsymbol{\Sigma}} \equiv {\bm{\Sigma}}^R$ and drop the superscripts for convenience. By introducing the non-interacting Green's function:
\begin{eqnarray}
	 {\mathbf G}_0^{-1}(\omega, t) = (\omega + \mu \pm i 0) \,\mathbf I - \mathbf h(t),
\end{eqnarray}
where $\pm i0$ is for the retarded and advanced function, respectively, the zeroth-order solution Eq.~(\ref{eq:G_RA_sol0}) is equivalent to the Dyson's equation parametrized by the central time $t$:
\begin{eqnarray}
	\label{eq:r-dyson}
	{\mathbf G}^{-1}(\omega, t) = {\mathbf G}_0^{-1}(\omega, t) - {\bm\Sigma}(\omega, t).
\end{eqnarray}
To solve the Green's functions, one needs to relate the self-energy to the full Green's functions, i.e. ${\bm\Sigma} = {\bm \Sigma}[ {\mathbf G} ]$, which is often intractable for general many-body interacting systems.

The central idea of DMFT is the approximation of a local self-energy, which, in the case of translation-invariant system, assumes that the electron self-energy is momentum-independent $\Sigma(\omega, \mathbf k) \approx \Sigma(\omega)$. This locality approximation is shown to be exact in the limit of infinite dimensions $d \to \infty$~\cite{metzner87,metzner89,hartmann89}, although DMFT often gives rather accurate results already for $d = 3$. Importantly, DMFT can be generalized to inhomogeneous systems by allowing a site-dependent, but still local, self-energy. In this approach, often called the real-space or statistical DMFT~\cite{potthoff99,ishida09,freericks04,tran07,okamoto08,helmes08,snoek08,gorelik10}, the interacting many-body problem is mapped to a set of quantum impurity models, one for each atom, which are solved self-consistently.  The real-space DMFT methods have been used to study Mott transitions in, e.g. Anderson-Hubbard model, correlated cold-atom systems, and heterostructures of correlated materials. Importantly, our new QMD scheme is also build on the real-space DMFT in order to treat random atom configurations in a liquid metal.

As discussed in the Introduction, the central idea of DMFT is the local self-energy approximation, which means that the self-energy matrix $\bm\Sigma$ in the Dyson equation~(\ref{eq:r-dyson}) is diagonal in atom-indices
\begin{eqnarray}
	\label{eq:local-Sigma}
	{\Sigma}_{i \sigma, j \sigma'}(\omega, t) = \delta_{ij} \delta_{\sigma \sigma'} {\Sigma}_{ii}(\omega, t).
\end{eqnarray}
For simplicity, here we have further assumed non-magnetic solutions, i.e. a spin-independent self-energy, for MD simulations.  The self-energy matrix is now diagonal in the atom-spin basis.  Based on this locality approximation, it is further assumed that the on-site self-energy $\Sigma_{ii}$ is obtained from solution of an atom-dependent quantum impurity model with the action
\begin{eqnarray}
	\label{eq:S}
	& & S_{\rm eff}^{(i)}(t) = U \int_0^\beta n_\uparrow(\tau) n_\downarrow(\tau) \\
	& & \quad -\sum_{\sigma}  \int_0^\beta d\tau \int_0^\beta d\tau' c^\dagger_\sigma(\tau)\, \bigl[\mathcal{G}^{(i)}_{0}(\tau - \tau'; \, t) \bigr]^{-1} \,c^{\,}_{\sigma}(\tau'). \nonumber
\end{eqnarray}
Here ${\mathcal{G}}^{(i)}_{0}(\omega; t) = \omega + \mu - \Delta_i(\omega; t)$ is an effective single-electron Green's function and $\Delta_i(\omega; t)$ is a time-varying hybridization to a fictitious bath that contains information about other atoms in the system. The subscript `0' here emphasizes that $\mathcal{G}_0$ is a ``bare" Green's function in absence of the Hubbard $U$ term. Notably, $\mathcal{G}_{0}$ plays a role similar to the Weiss field in the conventional static mean-field theory.  The full imaginary-time Green's function of the quantum impurity model is formally given by the expression  
\begin{eqnarray}
	\label{eq:G_imp}
	{G}^{(i)}_{\rm imp}(\tau, t) = \frac{-1}{\mathcal{Z}(t)}{\rm Tr}\bigl[ \mathcal{T}_{\tau} e^{S^{(i)}_{\rm eff}(t)} c_{\sigma}^\dagger(\tau) c^{\,}_{\sigma}(0) \bigr],
\end{eqnarray}
where $\mathcal{T}_{\tau}$ is the time-ordering operator, and $\mathcal{Z}$ is the partition function of the action. The real-frequency retarded Green's function $G^{(i)}_{\rm imp}(\omega, t)$ can then be obtained through analytical continuation of the corresponding Matsubara Green's function. 
The self-energy of this local impurity problem is computed from the Dyson equation
\begin{eqnarray}
	\label{eq:Sigma_imp}
	{\Sigma}^{(i)}_{\rm imp}(\omega, t) = {\mathcal{G}}^{(i)}_0(\omega, t)^{-1} - {G}^{(i)}_{\rm imp}(\omega, t)^{-1}. \quad
\end{eqnarray} 
This impurity self-energy is to be identified as the local self-energy given by the diagonal elements of the atomic self-energy matrix $\bm\Sigma$ in Eq.~(\ref{eq:local-Sigma}). 
\begin{eqnarray}
	\label{eq:Sigma-id}
	\Sigma_{ii}(\omega, t) = \Sigma^{(i)}_{\rm imp}(\omega, t).
\end{eqnarray}
The self-consistent condition of DMFT requires that the local Green's function coincides with the diagonal part of the system Green's function matrix:
\begin{eqnarray}
	\label{eq:dmft-eq}
	G_{i\sigma, i\sigma'}(\omega, t) = \delta_{\sigma\sigma'} \, G^{(i)}_{\rm imp}(\omega, t) .
\end{eqnarray}
It is worth noting that one can also think of the action in Eq.~(\ref{eq:S}) as a single fermion $c_\sigma$ couples to a bath of free fermions, from which the Weiss field $\mathcal{G}_0$ is generated. This is precisely the single-impurity Anderson model (SIAM) of a magnetic impurity hybridized with a conduction band. The two self-consistency equations~(\ref{eq:Sigma-id}), (\ref{eq:dmft-eq}) along with the two Dyson equations~(\ref{eq:r-dyson}), (\ref{eq:Sigma_imp}) essentially provide the following functional dependence for the local self-energy
\begin{eqnarray}
	\Sigma_{ii}(\omega, t) = \Sigma_{\rm SIAM}\left[ G_{ii}(\omega, t) \right].
\end{eqnarray}
which is an exact relation in the infinite dimension limit. 

In practical implementations of the real-space DFMT, the self-consistency is achieved through iterations; see Fig.~1 for details. One of the challenging part is the solution of the Anderson impurity model, formally given by Eq.~(\ref{eq:G_imp}). Several numerical techniques have been developed to solve the quantum impurity problem. One of the most powerful and widely used impurity solver is the continuous-time quantum Monte-Carlo algorithm~\cite{gull11}, which in principle can provide numerical exact results for the impurity problem. Other numerically exact impurity solvers include exact diagonalization, numerical renormalization group, and density matrix renormalization group~\cite{bula08}. Some of these solvers have been employed in real-space DMFT calculations for inhomogeneous Hubbard models [ ]. However, the huge overhead of these computationally sophisticated methods renders them difficult for large-scale dynamical simulations, where the impurity problems have to be solved for every atom at every time-step.

As a proof of principle, as well as to explore novel Mott physics in a liquid system, we adopt a modified version of the iterative perturbation theory (IPT)~\cite{georges92,kajueter96} as the local impurity solver, which allows for electron density away from the half-filling. Moreover, the calculation can also be directly performed on the real-time/frequency axes, without the need of analytical continuation. Although IPT is an approximate impurity solver, it often gives qualitatively correct results. Within IPT, the impurity self-energy  can be written in closed form~\cite{kajueter96}
\begin{eqnarray}
	\Sigma^{(i)}_{\rm imp}(\omega, t) = U n_i(t) + \frac{A_i(t) \, {\Sigma}^{(2)}_i\!(\omega, t)}{1 - B_i(t)\, \Sigma_i^{(2)}\!(\omega, t)}
\end{eqnarray}
where $\Sigma_i^{(2)}(\omega, t)$ is the second-order perturbation contribution to the self-energy~\cite{hartmann89}, and the coefficients $A_i$ and $B_i$ are determined in order to reproduce the correct moments of local density of states and the large-$U$ limit. Importantly, despite its appearance as a perturbation expansion, the IPT should be understood as an interpolation scheme as it gives the correct results both in the itinerant and atomic limit. 

A significant overhead of inhomogeneous DMFT method is numerical solution of the quantum impurity problem, which has to be solved for every atoms individually, and repeated several times until convergence at every time-step. Although the computational cost grows extremely fast when the number of site increase,  this step can be accelerated using parallel computation. Another computationally intensive step is solving the Dyson equation~(\ref{eq:r-dyson}) for the system Green's function, which requires computing the inversion of a sparse matrix $\mathbf M = [\mathbf G_0^{-1}(\omega, t) - \bm\Sigma(\omega, t)]$. The computation at different frequencies $\omega$ can again be parallelized. However, the bottleneck comes from the matrix inversion, whose complexity scales cubically with the system size when conventional linear-algebra algorithms are used. For convenience, the standard linear algebra package was used in this work for the matrix inversion. We note in passing that linear-scaling approximation methods  have recently been developed for solving the real-space Dyson equation~\cite{tang11,carrier11}.



\section{DMFT-MD simulation of Mott transitions}

\label{sec:results}

We perform the DMFT-MD simulations on the Hubbard liquid model~(\ref{eq:H}) based on the canonical NVT ensemble, i.e. with constant number of atoms ($N_a = 50 \sim 100$) in a fixed volume ($V$) and a constant temperature $T$ through the Langevin thermostat. Instead of volume $V$, it is convenient to introduce a parameter that characterizes the average inter-atomic distance  $r_s = (3 V / 4 \pi N)^{1/3}$. For given model parameters,  we introduce a reference energy $W_0$ defined as the average energy (per atom) of electron in absence of the Hubbard repulsion
\begin{eqnarray}
	\label{eq:W0}
	W_0 =- \frac{1}{N} \overline{ \sum_{ij, \sigma} h(R_{ij}) \bigl\langle c^\dagger_{i\sigma} c^{\,}_{j \sigma} \bigr\rangle_{U = 0}},
\end{eqnarray}
where the overline indicates MD averages.  The number of electrons is set at half-filling $N_e = N$ by tuning the chemical potential $\mu$.
All MD simulations presented below started with a random atomic configuration, and measurements were performed after the system reached the equilibrium after an initial transient period, the duration of which depends on, e.g. the temperature, the atomic density, and the Hubbard $U$ parameter. 

To characterize the Mott transitions, various quantities are measured and analyzed in the MD simulations. As customary in TBMD simulations, we divide the total energy of the liquid into three parts: the kinetic energy of atoms, the classical (empirical) repulsive potential, and the electronic energy, defined as follows 
\begin{eqnarray}
	& & E_{\rm kin} =   \sum_i \frac{\mathbf P_i^2}{2M}, \qquad \nonumber
	 E_{\rm rep} =  \frac{1}{2} \sum_{i \neq j} \phi(|\mathbf R_i - \mathbf R_j|), \\
	& & E_{\rm bind} =  \sum_{ij, \sigma} h_{ij} \langle c^\dagger_{i \sigma} c^{\,}_{j \sigma} \rangle + U \sum_i \langle n_{i\uparrow} n_{i \downarrow} \rangle. \quad
\end{eqnarray}
The correlation-induced localization of electrons is probed by computing the local double-occupancy, which is defined as the probability that two electrons with opposite spins occupy the same atom, and is given by the following expression~\cite{vilk97}
\begin{eqnarray}
	d_i = \langle n_{i\uparrow} n_{i\downarrow} \rangle = \frac{T}{U} \sum_m e^{-i \omega_m 0^+} \Sigma_{ii}(i \omega_m) G_{ii}(i\omega_m), \quad
\end{eqnarray}
where $\Sigma_{ii}(i\omega_m)$ and $G_{ii}(i\omega_m)$ are the on-site self-energy and Green's function at the Matsubara frequency $\omega_m$. A complementary measurement of the electron correlation effects is the local quasi-particle weight or $Z$-factor
\begin{eqnarray}
	Z_i^{-1} = 1 - \frac{\partial\, {\rm Re} \Sigma_{ii}(\omega+i 0^+)}{\partial \omega} \bigg|_{\omega = 0}
\end{eqnarray}
In Fermi-liquid theory, the averaged quasi-particle weight is related to the discontinuity of the Fermi distribution at the chemical potential. The inverse of $Z$ is also proportional to the enhancement of the effective mass of quasi-particles. The vanishing of the quasi-particle weight thus signals the breakdown of Fermi liquid behavior. Other electronic properties, such as the electrical and thermal conductivities, can be obtained from the electron Green's functions, which are readily computed at every MD-step.

The atomic configuration is best described by the radial pair distribution function $g(R)$ which gives the probability of finding another particle at a distance $R$ from an atom at the origin. Numerically, the distribution function is computed from the atomic pair distances subject to periodic boundary conditions
\begin{eqnarray}
	g(R) = \frac{1}{4\pi R^2 N \rho} \left\langle \sum_{i=1}^N \sum_{j \neq i}^N \delta(|\mathbf R_{ij} - R|) \right \rangle
\end{eqnarray} 
The distribution function approaches the limit $g(R) \to 1$ at large distances and vanishes for very small $R$, indicating the strong repulsive interaction at short distances. The atomic short-range correlation is encoded in the oscillatory behavior of the pair distribution function.

Dynamical behaviors of the atoms can be probed through various time-domain correlation functions computed from the atomic trajectories. For example, the self-diffusion coefficient $D$ is computed from the integral of the velocity auto-correlation function
\begin{eqnarray}
	\label{eq:diffusion}
	D = \frac{1}{3N} \sum_{i=1}^N \int_0^\infty \langle \mathbf v_i(t) \cdot \mathbf v_i(0) \rangle,
\end{eqnarray}  
It is worth emphasizing that the DMFT-MD method provides a unique tool to study the intriguing interplay between the electron correlation (characterized by the Hubbard repulsion) and atomic dynamics in metallic fluids. 

As discussed in Sec.~\ref{sec:intro}, here we consider two types of atomic liquids which are qualitatively distinguished by the size of the repulsive core $r_{\rm core}$ relative to the range of electron hopping $\xi$. These two characteristic lengths can be defined as $r_{\rm core} = \int_0^\infty \phi(R) R dR / \int_0^{\infty} \phi(R) dR$ and $\xi = \int_0^\infty h(R) R dR / \int_0^\infty h(R) dR$. For atomic liquids where these two lengths are of similar order, the electronic binding forces give rise to the formation of diatomic molecules, or dimers, in the ``conducting" phase. The resultant molecular liquid provides a simple model for liquid-phase hydrogens at high temperatures. On the other hand, for atoms with a much smaller repulsive core, the electron-mediated cohesive forces create large connected clusters that are essentially metallic. Mott transitions in this model system thus offers a canonical scenario for correlation-induced metal-insulator transition in an atomic liquid.

\subsection{Liquid-state Mott transition}

We first present DMFT-MD simulations of Mott transitions in an atomic liquid characterized by a short-ranged repulsive core $r_{\rm core} \ll \xi$. Specifically, we used an exponential decaying hopping function $h(R) = h_0 \exp(-R/ \xi)$ where $h_0$ and $\xi$ define the energy and length scales of the model, respectively. The parameters for the pair potential $\phi(R) = \phi_0 \exp[-R / \ell - b (R/\ell)^4]$ are $\ell = 0.86 \xi$ and $b = 0.1$.   The number of atoms in the following simulations is $N = 50$ with an atomic density specified by an average distance $r_s = 2.26 \xi$. For convenience, the energy is to be measured with respect to the non-interacting electronic energy $W_0$ defined in Eq.~(\ref{eq:W0}), while lengths are measured in terms of the equilibrium distance $R_0$ of the binding curve $E(R) = -2 h(R) + \phi(R)$ of an atomic pair.

\begin{figure}[t]
\includegraphics[width=\linewidth]{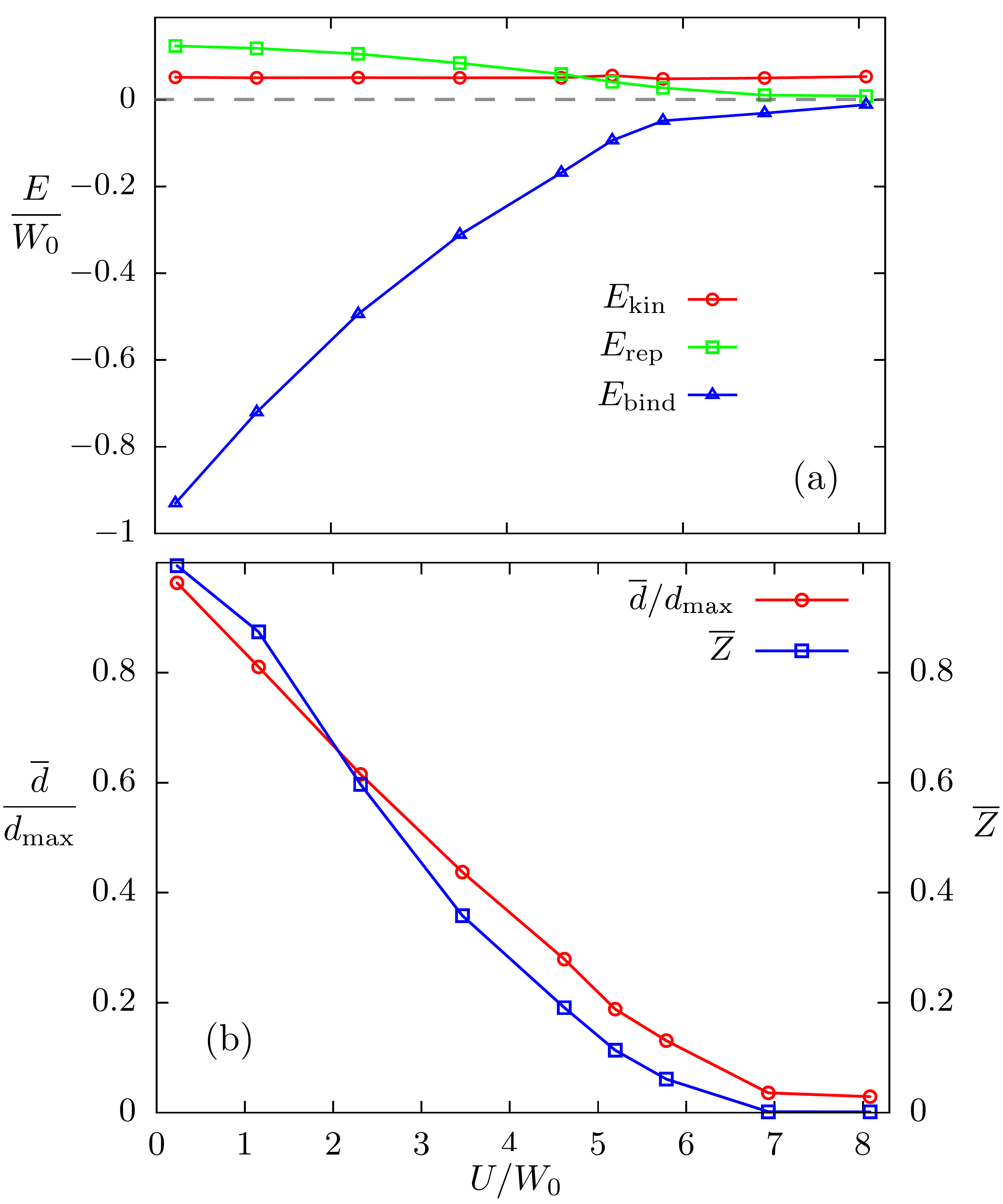}
\caption{(a) The various energy densities of the Hubbard liquid model versus the on-site Coulomb repulsion~$U$. (b) the averaged double occupancy $\overline{d} =  \sum_i d_i / N$ and averaged quasi-particle weight $\overline{Z} =  \sum_i Z_i / N$ as a function of on-site Coulomb interaction $U$. The double-occupancy is normalized to its maximum value $d_{\rm max}=\langle n_\uparrow \rangle \langle n_\downarrow\rangle = 0.25$ in the uncorrelated state at half-filling.}
\label{fig:mott-trans1}
\end{figure}

Following previous works on Hubbard models, we consider the metal-insulator transition driven by increasing the strength of on-site repulsion $U$ relative to the bandwidth $W_0$ of electron hopping. Fig.~\ref{fig:mott-trans1}(a) shows the dependence of various energy components of the  Hubbard liquid model on the on-site Coulomb repulsion~$U/W_0$. 
Because of the Langevin thermostat which directly couples to the ionic motion, the kinetic energy remains roughly a constant $E_{\rm kin} / N \approx 3 k_B T /2$ that is proportional the temperature which is set at $T = 0.03 W_0$. On the other hand, the magnitude of the electronic binding energy $|E_{\rm bind}|$ decreases monotonically with increasing ratio $U/W_0$. At small Hubbard $U \sim 0$, the electronic energy is roughly $E_{\rm bind} \sim -  N W_0$, consistent with our definition of $W_0$ as a measure of the binding energy per atom. The electronic energy approaches zero for~$U > U_c \sim 7 W_0$. The energy of the repulsive pair interaction, $E_{\rm rep}$, is correlated to the electronic energy and decreases steadily as~$U$ approaches the critical value. 

The loss of the binding energy at $U > U_c$ indicates the disappearance of chemical bonding that provides the attractive force between atoms. Indeed, the concomitant decrease of the repulsive interaction $E_{\rm rep}$ also means that the average separation between atoms is enlarged.   This in turn implies the suppression of electron hopping, and a tendency toward localization. This is indeed confirmed by the $U$-dependence of the double-occupancy $\overline{d}$, averaged over all atoms in the liquid, shown in Fig.~\ref{fig:mott-trans1}(b). The average double-occupancy $\overline{d}$ decreases monotonically as the Hubbard $U$ approaches the critical value. 

Since in average there is one electron per atom at half-filling, the reduced double-occupancy means suppressed electron hopping. This in turn indicates the suppression of quasi-particle excitations, which can be either interpreted as the enhancement of effective mass or the loss of quasiparticle weight. Our simulations show that the averaged quasiparticle weight $\overline{Z} = \sum_i Z_i / N_a$ tends to zero above the critical $U_c$. The results summarized in Figs.~\ref{fig:mott-trans1} clearly show an electronic phase transition that is due to correlation-induced electron localization in the Hubbard liquid model.

\begin{figure}
\includegraphics[width=0.9\linewidth]{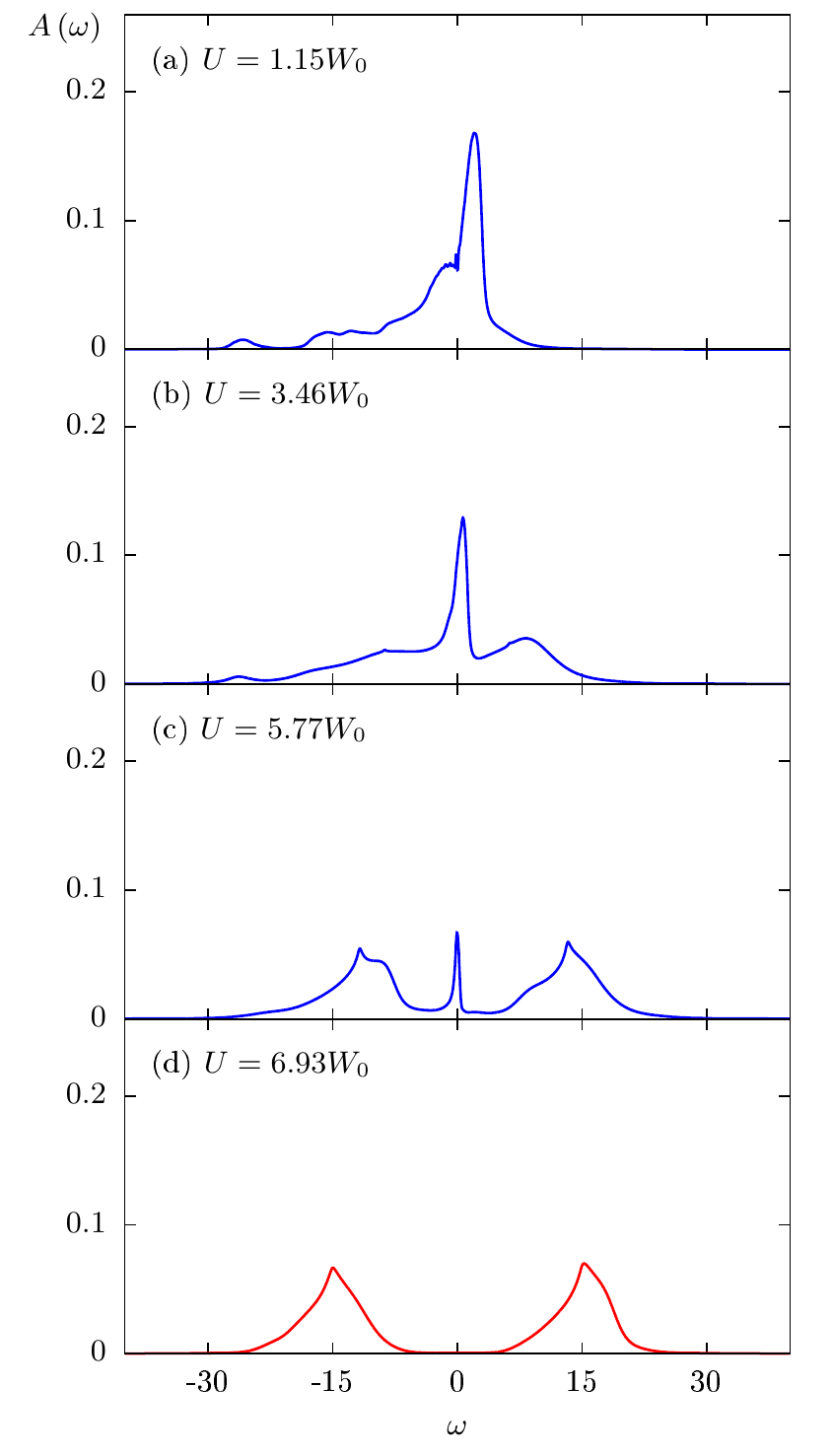}
\caption{The spatially and temporally averaged electron spectral function $A(\omega)$ at varying values of Hubbard parameter obtained from DMFT-MD simulations. The Mott transition occurs at a critical $U_c \sim 6.9 W_0$. Panel~(c) shows a three-peak structure consisting of two Hubbard bands and a sharper quasi-particle peak at the center.}
\label{fig:spectral1}
\end{figure}

As discussed in the Introduction, the DMFT approach allows us to investigate the evolution of the complex electronic excitations during a Mott transition, which is beyond the capability of self-consistent single-particle methods such as the Gutzwiller approximation. In the real-space DMFT formalism, a local spectral function can be computed from the imaginary part of the diagonal lesser Green's function:
\begin{eqnarray}
	A_i(\omega; t) = -\frac{1}{\pi} G^<_{ii}(\omega; t).
\end{eqnarray}
Here $t$ is the slowly varying central time introduced in Sec.~\ref{sec:adiabatic}. A spatially and temporally averaged spectral function $A(\omega) = \frac{1}{N\, M} \sum_{i=1}^N \sum_{k = 1}^M A_{ii}(\omega; t_k)$ can then be obtained by averaging over all atoms and over time sequences $\{ t_1, t_2, \cdots , t_M \}$ of the MD simulations. The resultant average spectral function is shown in Fig.~\ref{fig:spectral1} at various values of the Hubbard parameter. At small $U$, the spectral function shown in Fig.~\ref{fig:spectral1}(a) corresponds to the density of states (DOS), which is asymmetric with respect to the Fermi level $\omega = 0$, of the underlying metallic bonding of the atomic clusters.

Upon further increasing the Hubbard repulsion, a quasi-particle peak starts to develop at $\omega = 0$ on top of the spectrum of the tight-binding Hamiltonian, as shown in Figs.~\ref{fig:spectral1}(b) and (c). The narrowing of this central peak can be accounted for by the Fermi liquid theory. Close to the Mott transition point, the spectral function exhibits a characteristic three-peak structure as shown in Fig.~\ref{fig:spectral1}(c). In addition to the quasi-particle peak at the center, two Hubbard bands, which originates from atomic-like excitations, begin to emerge. Above the Mott transition $U > U_c$, strong electron correlation causes the transfer of the spectral weight from the low-energy peak to the high-frequency Hubbard bands, leading to the vanishing of the quasi-particle peak.

\begin{figure}[b]
\includegraphics[width=0.9\linewidth]{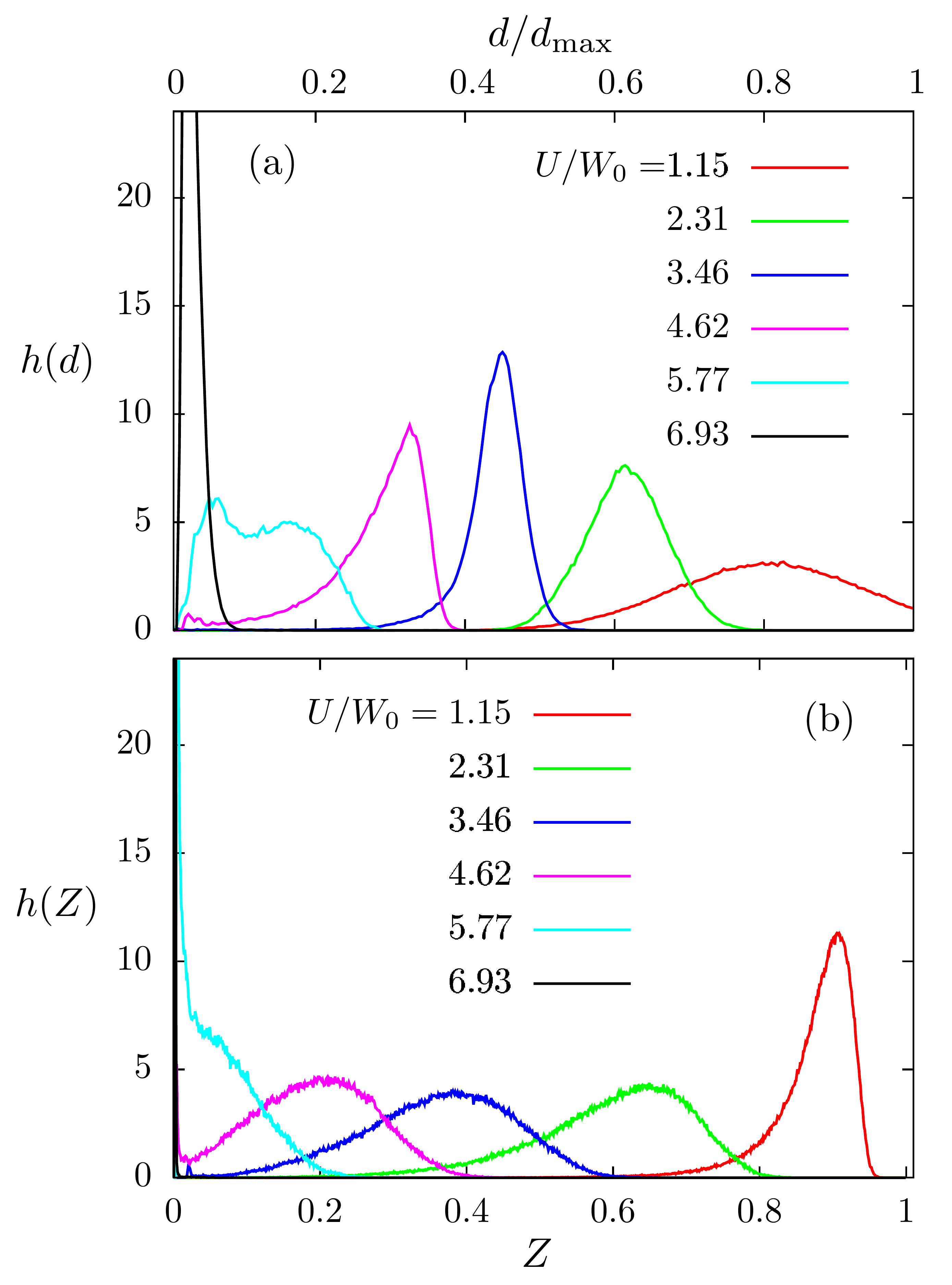}
\caption{Histogram of (a) local double occupancy $d_i$ and (b)~quasi-particle weight $Z_i$ at varying Hubbard repulsion obtained from the DMFT-MD simulations. The Mott transition occurs at a critical $U_c \sim 6.9 W_0$.}
\label{fig:hist1}
\end{figure}

The Mott-transition scenario in the Hubbard liquid model outlined above is overall similar to those observed in the crystalline and some amorphous systems, at least within the DMFT framework~[...]. However, the mobility of atoms in a liquid gives rise to a highly heterogeneous  and dynamical electronic state, especially in the vicinity of the critical point. To gain insight into the dynamical nature of the liquid-state Mott transition, Fig.~\ref{fig:hist1} shows the histograms of local double-occupancy $d_i$ and quasi-particle weight $Z_i$ at various Hubbard parameters. Both distributions $h(d)$ and $h(Z)$ show a single broad peak for most values of Hubbard $U \lesssim U_c$ on the metallic side. This result can be understood as the existence of one types of atoms with varying degree of electron delocalization. Both the average value and the peak-value decrease as the system approaches the Mott transition. Above the critical point $U > U_c$, the distribution $h(d)$ exhibits a sharp peak at a small yet nonzero $d_m \sim 0.005$, while the quasi-particle weight of all atoms becomes zero, consistent with the results shown in Fig.~\ref{fig:mott-trans1}(b).

\begin{figure}[b]
\includegraphics[width=0.99\linewidth]{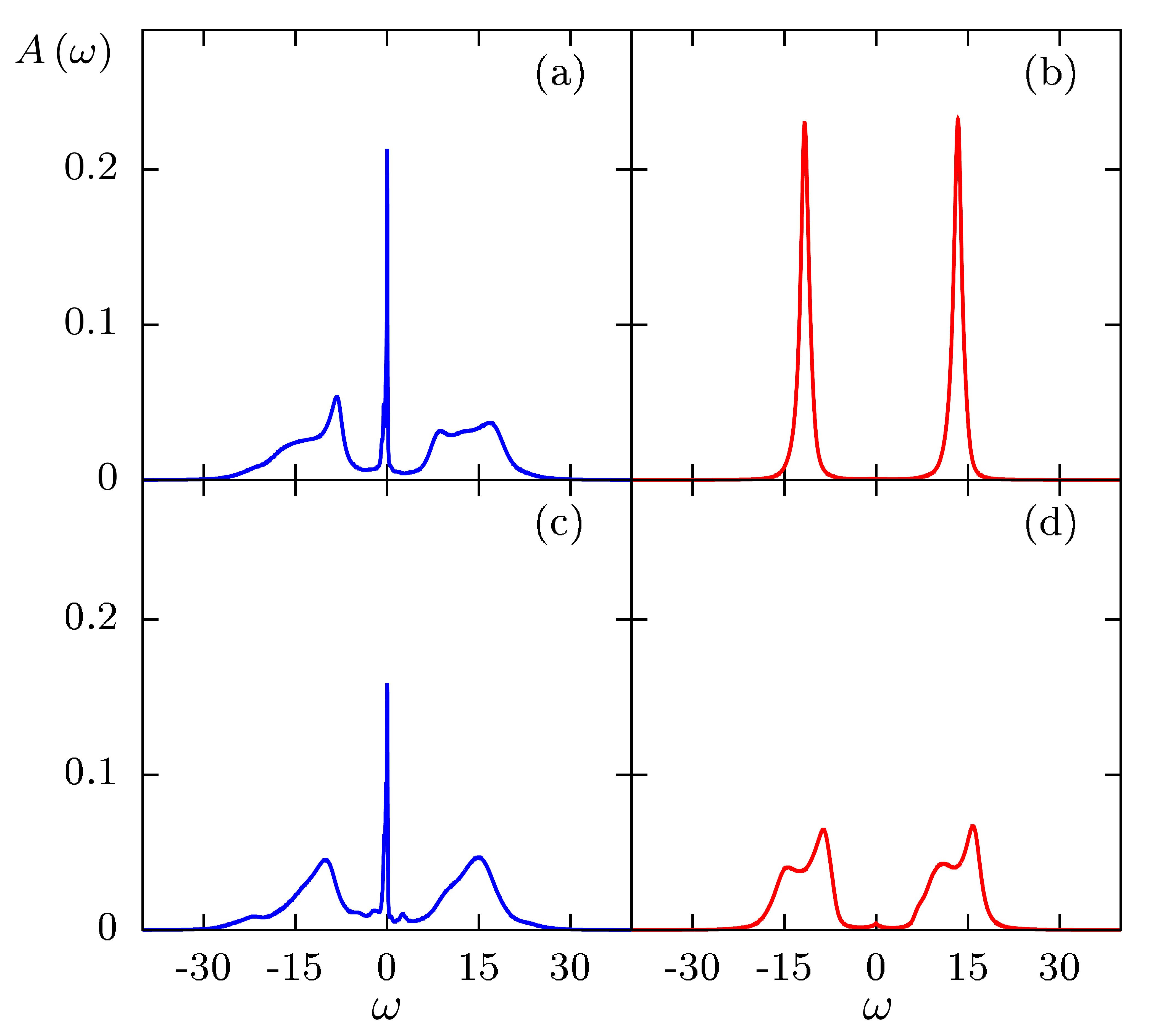}
\caption{Local spectral functions $A_i(\omega, t)$ on four different atoms at a given instant of the DMFT-MD simulations. The Hubbard parameter is $U = 5.77 W_0$.}
\label{fig:local-Aw1}
\end{figure}

Interestingly, in the vicinity of the Mott transition, the distribution of double-occupancy $h(d)$ exhibits a two-peak structure, e.g. at $U = 5.77 W_0$, on the metallic side. The corresponding distribution function $h(Z)$ can be viewed as a mixture of a delta-peak $\delta(Z)$ and a broad peak also centered at $Z = 0$. These results indicate the coexistence of two species of atoms: one group with a finite double-occupancy and quasi-particle weight, and the other type of atoms with localized electrons. The emergence of such heterogeneous electronic states can also be seen in the local spectral functions $A_i(\omega, t)$ shown in Fig.~\ref{fig:local-Aw1} for four randomly chosen atoms. At the given instant, the system exhibits two distinct types of local spectral functions. The persistence of the quasi-particle peak accompanied by the two Hubbard bands indicate a metallic atom with delocalized electrons. At the same time, for those atoms that undergo a local Mott transition, the spectral weight of the quasi-particle is completely transferred to the high-energy Hubbard bands. Finally, the shape of the Hubbard bands, which depends on the local atomic enviroment, varies from atom to atom, a further manifestation of the heterogeneity of the strongly correlated metallic state.

\begin{figure}
\includegraphics[width=0.95\linewidth]{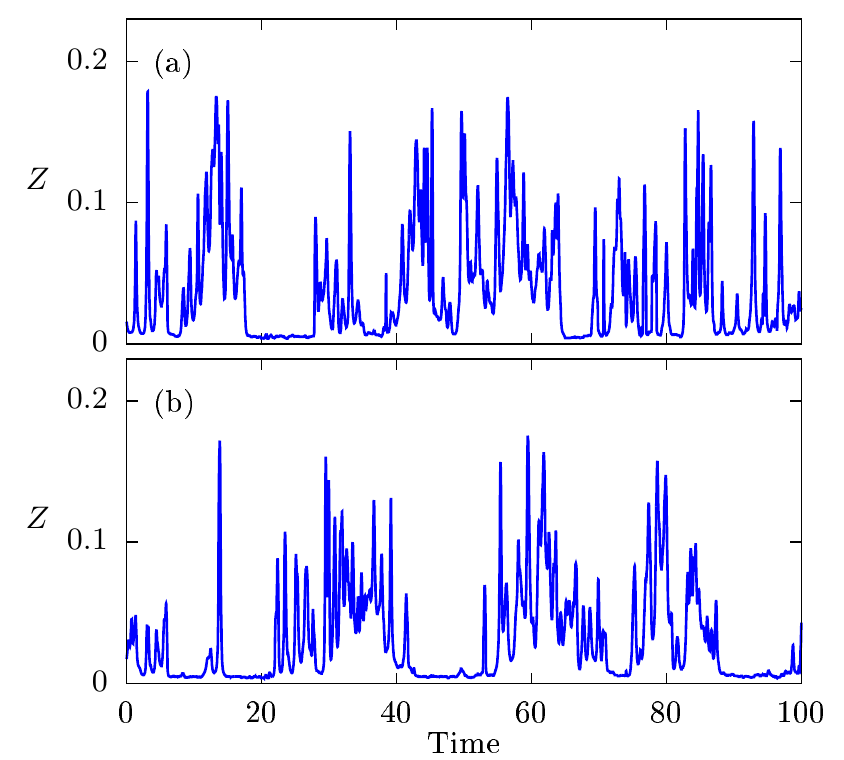}
\caption{Local quasi-particle weight $Z_i$ as a function of time for two randomly chosen atoms at $U = 5.77 W_0$.}
\label{fig:Z_vs_t1}
\end{figure}

It is also worth noting that the heterogeneity is highly dynamical and occurs at the atomic scale, contrary to the mesoscopic phase-separated states with multiple quasi-static competing phases. To see this, we plot in Fig.~\ref{fig:Z_vs_t1} the time dependence of local quasi-particle weight $Z_i(t)$ for two randomly chosen atoms obtained from the DMFT-MD simulations at $U = 5.77 W_0$. Interestingly, both time traces exhibit windows of vanishing $Z$ interspersed between periods of nonzero and fluctuating quasi-particle weight, which means that the atom-specific Mott transition is not only local but also temporary in this highly dynamical state close to the Mott transition. Moreover, since atoms need to be embedded in a finite cluster in order to remain metallic, the fluctuating traces of $Z_i$ in Fig.~\ref{fig:Z_vs_t1} also highlight the dynamical nature of the atomic clusters: atoms constantly leave and rejoin the clusters via local intermittent Mott transitions.

\begin{figure}
\includegraphics[width=0.95\linewidth]{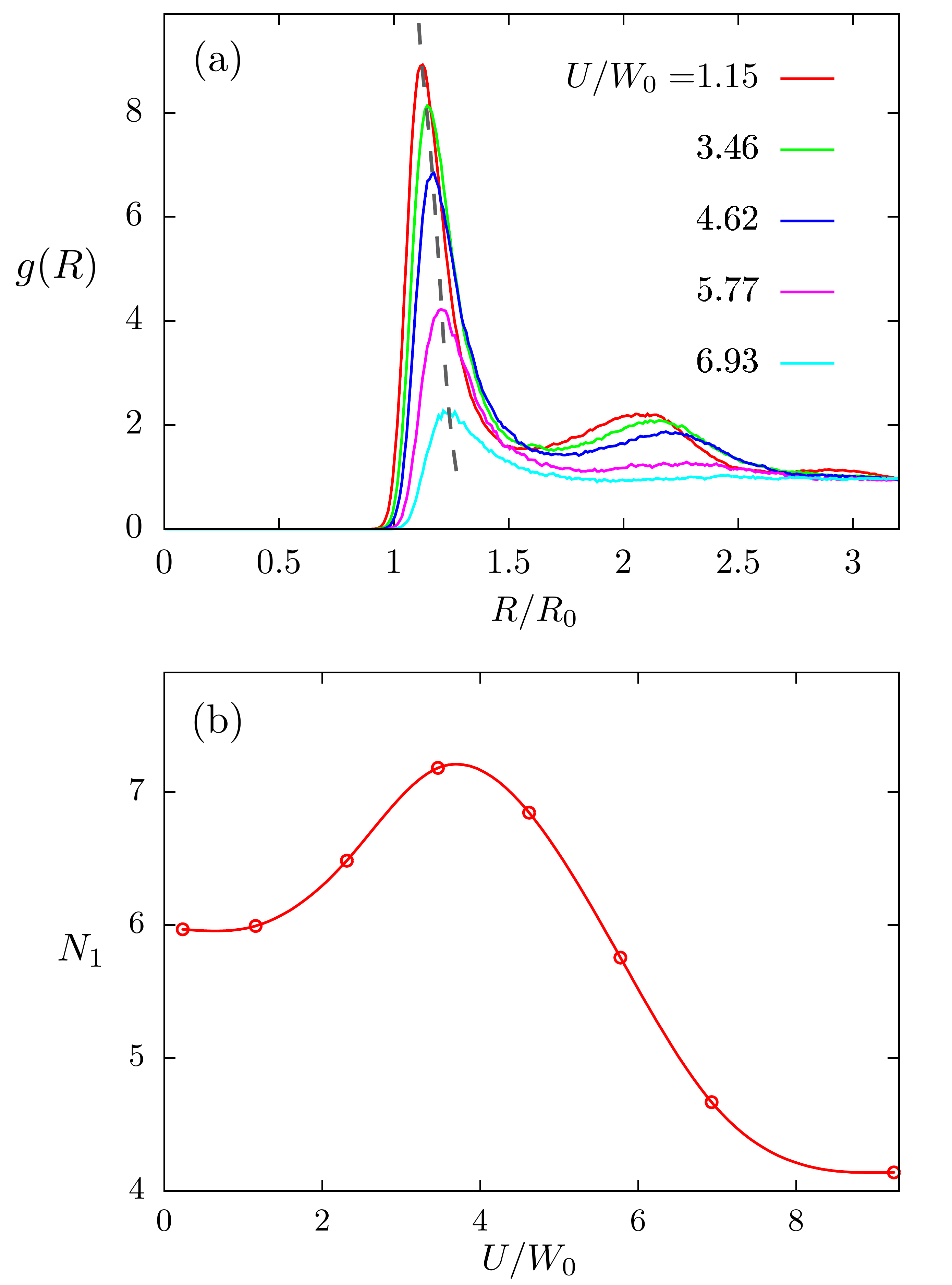}
\caption{(a) The pair distribution function $g(R)$ at different values of Hubbard parameter. The reference length $R_0$ is given by the equilibrium distance of the binding curve $E(R) = -2 h(R) + \phi(R)$ between a pair of atoms. The Mott transition occurs at $U_c \sim 7 W_0$. (b) the effective coordination number $N_1$ versus Hubbard parameter $U$ obtained from the DMFT-MD simulations. }
\label{fig:pair-dist1}
\end{figure}

To study the structure of these dynamical atomic clusters, we plot in Fig.~\ref{fig:pair-dist1}(a) the pair distribution function $g(R)$ at various Hubbard parameters obtained from the DMFT-MD simulations. The short-range atomic correlation is marked by a pronounced peak at $R^* \sim 1.1 R_0$, which shifts upward slightly with increasing $U$; here $R_0$ is the equilibrium distance of an isolated pair of atoms. At small $U$, the distribution function also shows a second smaller peak at a larger distance. Interestingly, these atomic clusters are not characterized by the kind of strong correlations that are typical of dense liquid. For example, other than the short-distance repulsive region defined by a radius of roughly $R_{\rm rep} = 0.98 R_0$, the atoms do not avoid each other at longer distances, as indicated by the fact that the value of $g(R)$ is always greater than unity for~$R > R_{\rm rep}$.

The two peaks originate from longer-range attractive interactions between atoms. To quantify the short-range atomic correlation, we compute the effective coordinate number $N_1$, which can be viewed roughly as the number of the ``nearest-neighbors". It is defined as the integral: $N_1 = \rho_0 \int_0^{R_m} g(R) 4\pi R^2 dR$, where $\rho_0$ is the atomic number density and $R_m$ is the first minimum of the the distribution function. The dependence of this coordination number on the Hubbard repulsion is shown in Fig.~\ref{fig:pair-dist1}(b). Interestingly, the effective coordination is enhanced at intermediate $U$ before dropping to a smaller value in the Mott phase. Although the average distance of these effective nearest neighbors also increases slightly with increasing $U$, this non-monotonic behavior implies a strengthened atomic connection, at least structurally, induced by the electron correlation.

Before closing this subsection, we note that the DMFT-MD simulations here give the overall picture of correlation-induced metal-insulator transition in a generic metallic fluid as originally envisioned by Mott. This new scenario of Mott transition in an atomic liquid should share some similarities with those in crystalline and amorphous solids. For example, electronically the Mott transitions in all cases are characterized by a spectral function with a three-peak structure and the transfer of spectral weight from the central quasi-particle peak to the high-frequency Hubbard bands. On the other hand, the liquid-state case, even with the Born-Openheimer adiabatic approximation, is more than an ensemble of disordered atoms. The dynamically fluctuating clusters and local temporary Mott transitions in the strongly correlated metallic state of the liquid represents a new state of matter that is unveiled by the DMFT-MD simulations. The persistence of the conducting atomic clusters also has important implications for the density-driven MIT in expanding fluid metals, to be discussed in Sec.~\ref{sec:discussion}.


\subsection{Mott transition and dissociation of dimers}

The mobility of atoms in a liquid adds a new dimension to the physics of Mott transitions. Indeed, as highlighted by the examples of dynamical atomic clusters, the intriguing interplay between chemical bonding and electron correlation could lead to the appearance of quasi-stable atomic structures and fundamentally change the nature of metal-insulator transitions. Here we demonstrate such a scenario in this correlated $s$-band liquid model. We consider a TBMD model with exponential decaying hopping and pair-potential functions. Again, the parameters $h_0$ and $\xi$ in the hopping coefficient $h(R) = h_0 \exp(-R / \xi)$ set the energy and length scales, respectively, of the system. And the parameters for the pair potential $\phi(R) = \phi_0 \exp(-R / \ell)$ are: $\phi_0 = 4 h_0$ and $\ell = 0.7 \xi$. The resultant electronic energy at $U = 0$ is $W_0 \sim 0.22 h_0$. We again performed NVT-type MD simulations with number of atoms $N = 50$ and temperature $T = 0.03 W_0$. The volume of the simulation box is fixed such that the atomic number density $r_s = 1.2 R_0$, where $R_0$ is again the equilibrium distance of an isolated pair of atoms. We again focused on the half-filled system with~$N_e = N$.

\begin{figure}
\includegraphics[width=\linewidth]{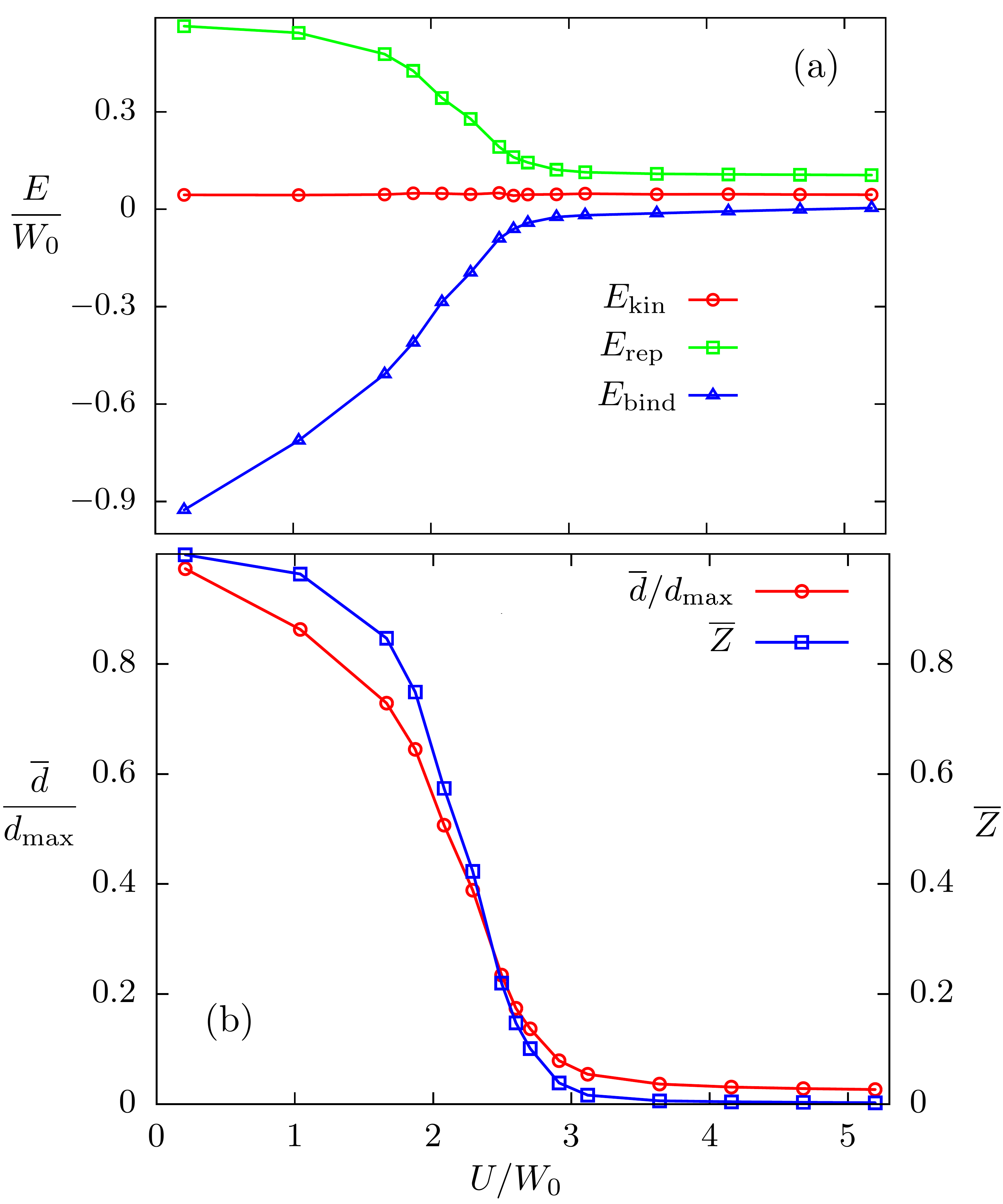}
\caption{(a) The various energy densities of the Hubbard liquid model with $r_{\rm core} \sim \xi$ versus the Hubbard parameter~$U$. (b)~the averaged double occupancy $\overline{d} =  \sum_i d_i / N$ and the averaged quasi-particle weight $\overline{Z} =  \sum_i Z_i / N$ as a function of $U$. The double-occupancy is normalized to its maximum value $d_{\rm max}=\langle n_\uparrow \rangle \langle n_\downarrow\rangle = 0.25$ in the uncorrelated state at half-filling.}
\label{fig:mott-trans2}
\end{figure}

Compared with the model system discussed in the previous section, the atomic liquid corresponding to these parameters is characterized by a relatively larger repulsive core such that $r_{\rm core} \sim \ell \sim \xi$.
The basic energetics of the Mott transition in this atomic liquid is summarized in Fig.~\ref{fig:mott-trans2}(a). The kinetic energy of atoms is well controlled by the Langevin thermostat, as indicated by the approximately constant curve of $E_{\rm kin}$. On the other hand, the magnitude of binding energy as well as the repulsive potential energy decrease with increasing Hubbard parameter; both tend to zero at $U \gtrsim U_c \approx 3.1 W_0$. The vanishing of electronic binding or cohesive energy at $U_c$ is again due to the localization of electrons in a Mott transition, as indicated by the $U$-dependence of the averaged double-occupancy $\overline{d}$ and quasi-particle weight $\overline{Z}$ shown in Fig.~\ref{fig:mott-trans2}(b). However, contrary to the almost linear decrease of both quantities in the previous case [see Fig.~\ref{fig:mott-trans1}(b)], there are two stages in the evolution of both quantities with the Hubbard parameter: a slower decrease in the weak correlation regime, followed by a faster reduction at intermediate $U$.

\begin{figure}
\includegraphics[width=0.9\linewidth]{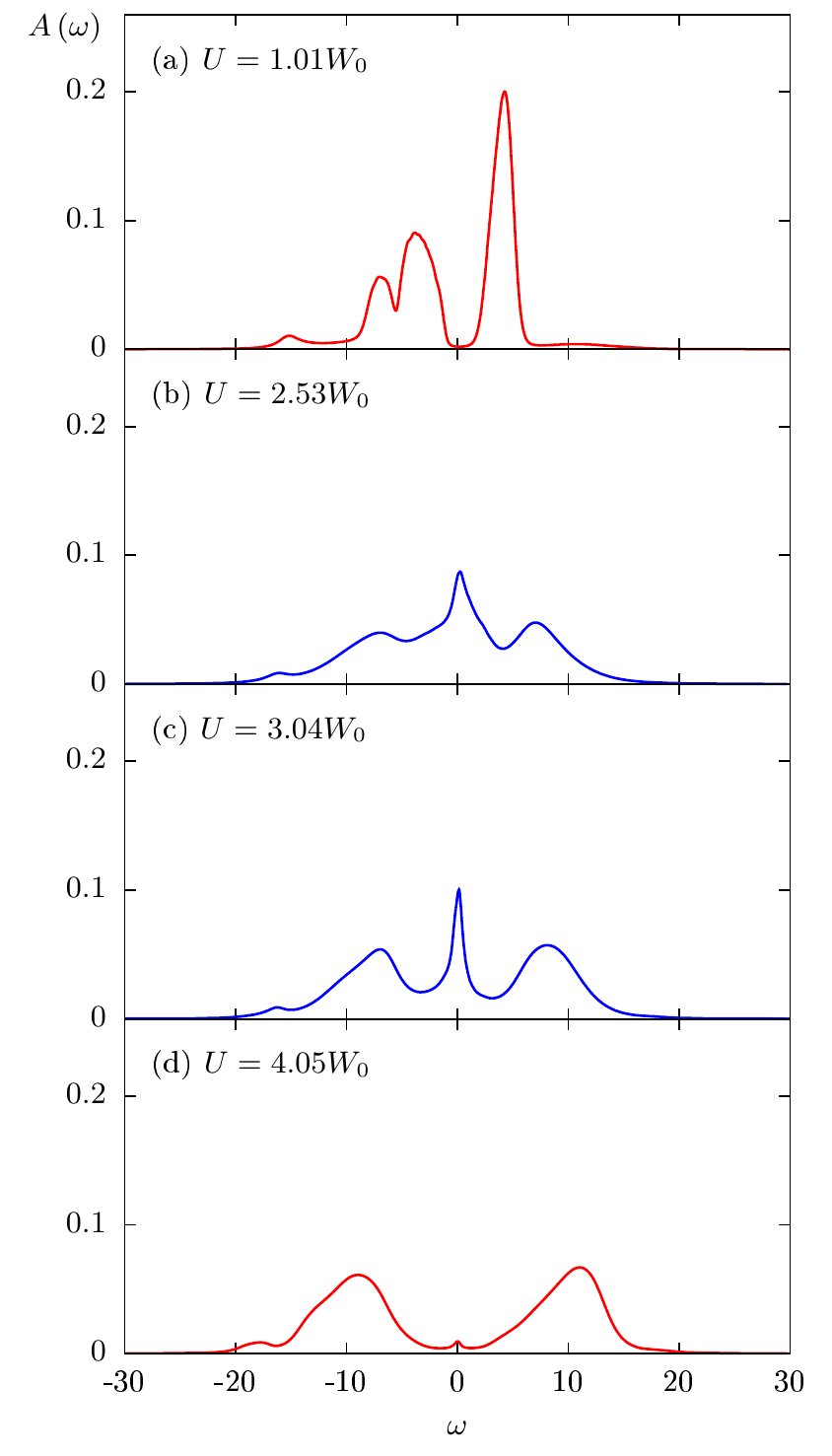}
\caption{The spatially and temporally averaged electron spectral function $A(\omega)$ at varying values of Hubbard parameter obtained from DMFT-MD simulations. The Mott transition occurs at a critical $U_c \sim 3.1 W_0$. The vanishing spectral weight at the Fermi level $\omega = 0$ in both panels~(a) and (d) indicate electronic insulator for both states at small $U$ and large~$U > U_c$.}
\label{fig:spectral2}
\end{figure}

These two distinct regimes, both in the ``conducting" phase of the Mott transition, are in fact two distinct states both electronically and structurally. To see this, we plot in Fig.~\ref{fig:spectral2} the spatially and temporally averaged spectral function $A(\omega)$ of the atomic system at various Hubbard parameters. Interestingly, a spectral gap can be seen at the Fermi level $\omega = 0$ even at $U = 1.01 W_0$ well below the critical $U_c$ of Mott transition. In fact, the spectrum at small $U$, which is highly asymmetric with respect to Fermi level, is similar to that of a band insulator. Physically, this result indicates the formation of diatomic quasi-molecules, or dimers, stabilized by a valence bond. The lower and higher bands thus correspond to the bonding and anti-bonding states of the molecules. Consequently, even though electrons are delocalized in this small $U$ regime, the electrons are confined to individual dimers, giving rise to the insulating nature of this liquid state. In fact, as the diatomic molecules are weakly coupled to each other, this state is better described as a molecular gas. 

As electron correlation is increased, the spectral gap between the bonding and anti-bonding states is closed and a quasi-particle peak develops in the intermediate-$U$ regime; see Figs.~\ref{fig:spectral2}(b) and (c). The gap-closing thus corresponds to the dissociation of dimers and the formation of a true metallic state. Moreover, a three-peak structure with two Hubbard bands enclosing a central quasi-particle peak starts to emerge as the system approaches the Mott transition. Above the critical $U_c$, only the two Hubbard bands remain in the Mott insulating state. The vanishing of quasi-particle weight also means the localization of electrons and loss of cohesive forces, or chemical bonds. As a result, the dominant interatomic interaction is the short-range repulsive interaction $\phi(R)$ in this insulating state. 

\begin{figure}
\includegraphics[width=0.95\linewidth]{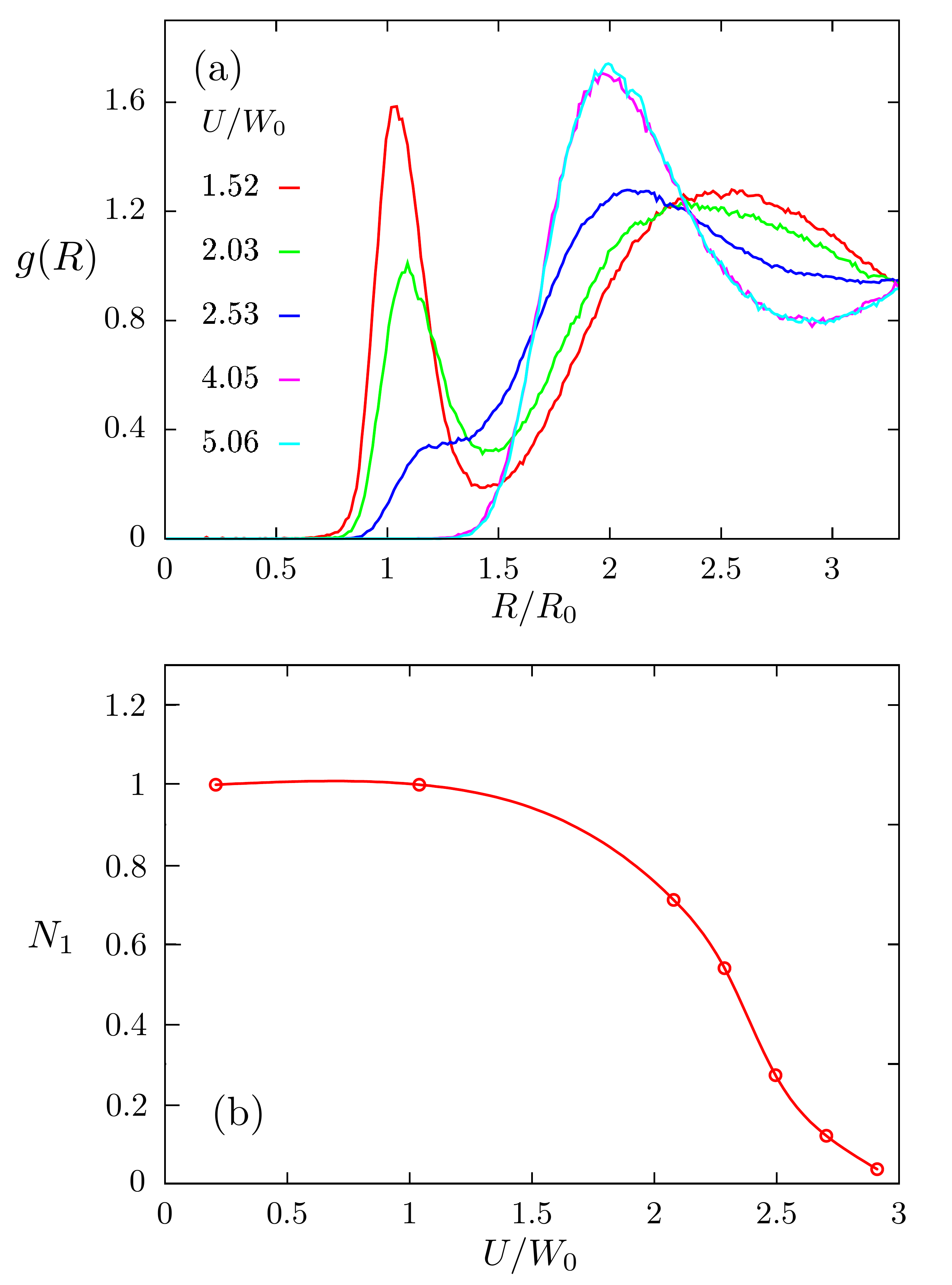}
\caption{(a) The pair distribution function $g(r)$ at different values of Hubbard parameter. The Mott transition occurs at $U_c \sim 3.1 W_0$. (b) the effective coordination number $N_1$ versus Hubbard parameter $U$ obtained from the DMFT-MD simulations. }
\label{fig:pair-dist2}
\end{figure}

The above scenario, namely a transition from an insulating molecular liquid to an insulating Mott liquid with an intermediate metallic state, is mainly inferred from the electronic spectral functions. This picture is also supported by the atomic structural results, such as the pair distribution function $g(R)$ shown in Fig.~\ref{fig:pair-dist2}(a). At small $U$, the distribution function exhibits a prominent peak at a distance $R \sim R_0$, where $R_0$ is the equilibrium distance of an isolated pair of atoms. This peak thus corresponds to the formation of dimers at the energy minimum of the binding curve $E_B(R) = -2 h(R) + \phi(R)$. Indeed, by integrating the pair distribution $g(R)$ from zero to its first minimum gives an effective coordination number $N_1 \sim 1$ at small $U$; see Fig.~\ref{fig:pair-dist2}(b), which means there is essentially only one nearest-neighbor atom in this weak correlation regime.

Upon increasing $U$, the enhanced electron correlation leads to the suppression of the dimer peak, which physically corresponds to the dissociation of the quasi-diatomic molecules. This process also manifests itself in the gradual reduction of the effective coordination number $N_1$ from one toward zero. Above the critical $U_c$, the dimer peak completely vanishes in the Mott phase, and the pair distribution function is now dominated by a pronounced peak at $R \sim 2 R_0$. The $g(R)$ at large $U > U_c$ essentially describes the short-range correlation that results from the repulsive pair potential.

\begin{figure}
\includegraphics[width=0.9\linewidth]{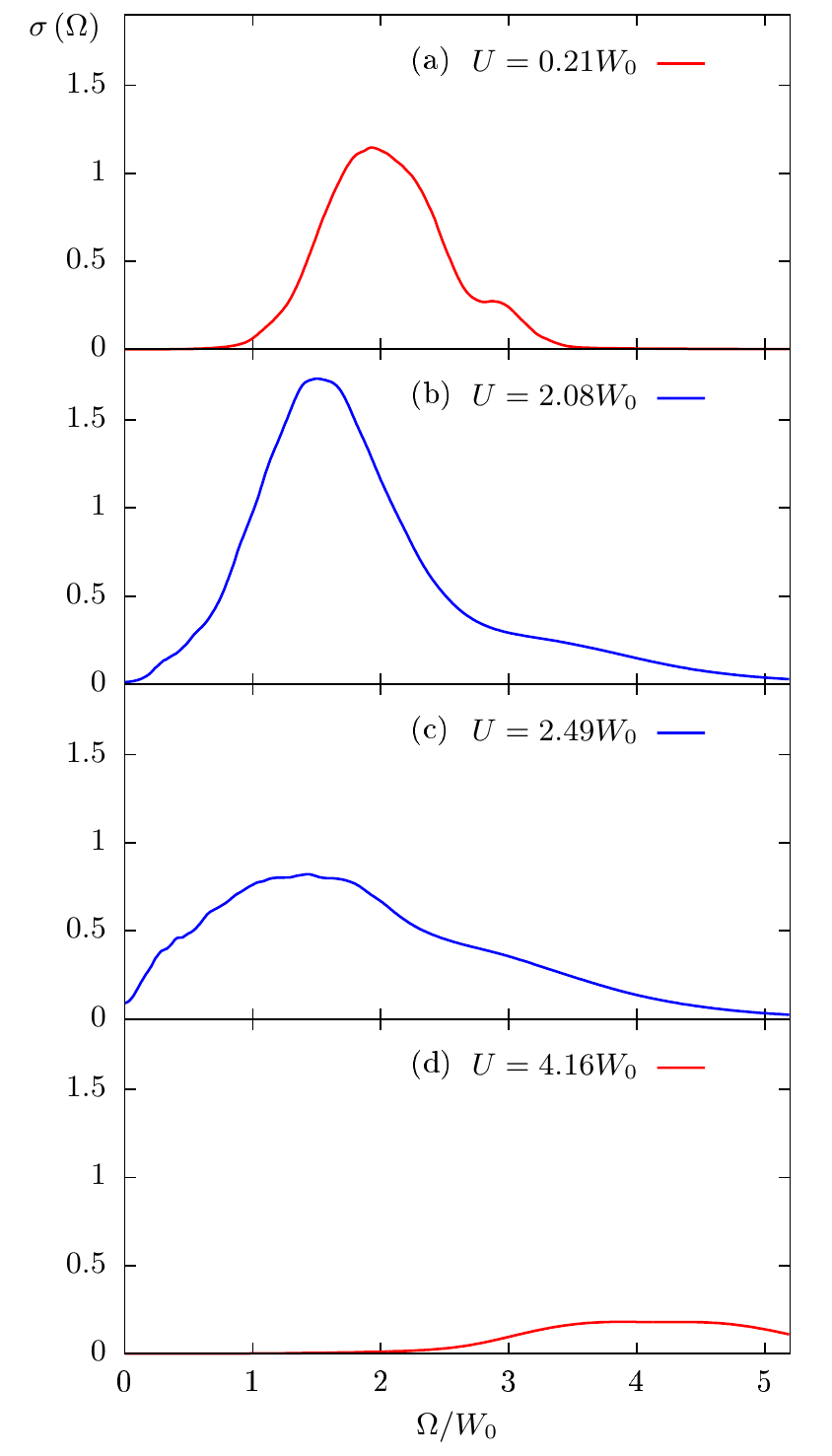}
\caption{Time averaged optical conductivity $\sigma(\Omega)$ as a function of frequency $\Omega$ obtained from the DMFT-MD simulations. The d.c. conductivity $\sigma(\Omega \to 0)$ becomes finite at $U \sim 2.1 W_0$, signaling the dissociation of dimers and the emergence of the intermediate metallic phase.}
\label{fig:conductivity}
\end{figure}

With the electron's Green's functions from the real-space DMFT, various electronic transport coefficients of the atomic liquid can be readily computed. Fig.~\ref{fig:conductivity} shows the optical conductivity $\sigma(\Omega)$ as a function of frequency~$\Omega$ at different values of Hubbard repulsion. In the small-$U$ insulating regime, the dc conductivity vanishes and the optical conductivity is marked by a broad peak corresponding to transitions from the bonding to the anti-bonding states of the diatomic molecules. The dissociation of these dimers at the intermediate region gives rise to a finite dc conductivity $\sigma(\Omega \to 0) > 0$; see Figs.~\ref{fig:conductivity}(b) and (c). Notably, a very broad band ranging from $\Omega =0$ up to $\Omega \sim 5 W_0$ emerges in the strongly correlated metallic states. As electron correlation is further increased above $U_c$, the low-frequency absorption disappears in the Mott insulating phase; the wide band at energies $\Omega \sim U$ in panel~(d) results from optical transitions between the two Hubbard bands.

The atomic liquid in the intermediate-$U$ regime is not only metallic, but also is a highly heterogeneous state. In fact, the dimers are only partially dissociated in this intermediate phase which can be viewed as a transitional regime connecting the insulating dimer phase to the Mott phase. Structurally, the mixture of bounded and dissociated dimers in the intermediate states can be seen in the $g(R)$ curve at $U = 2.53 W_0$ in Fig.~\ref{fig:pair-dist2}(a), where a residual dimer peak coexists with the broad second peak at $R \sim 2 R_0$ characterizing the Mott phase.

\begin{figure}[b]
\includegraphics[width=0.99\linewidth]{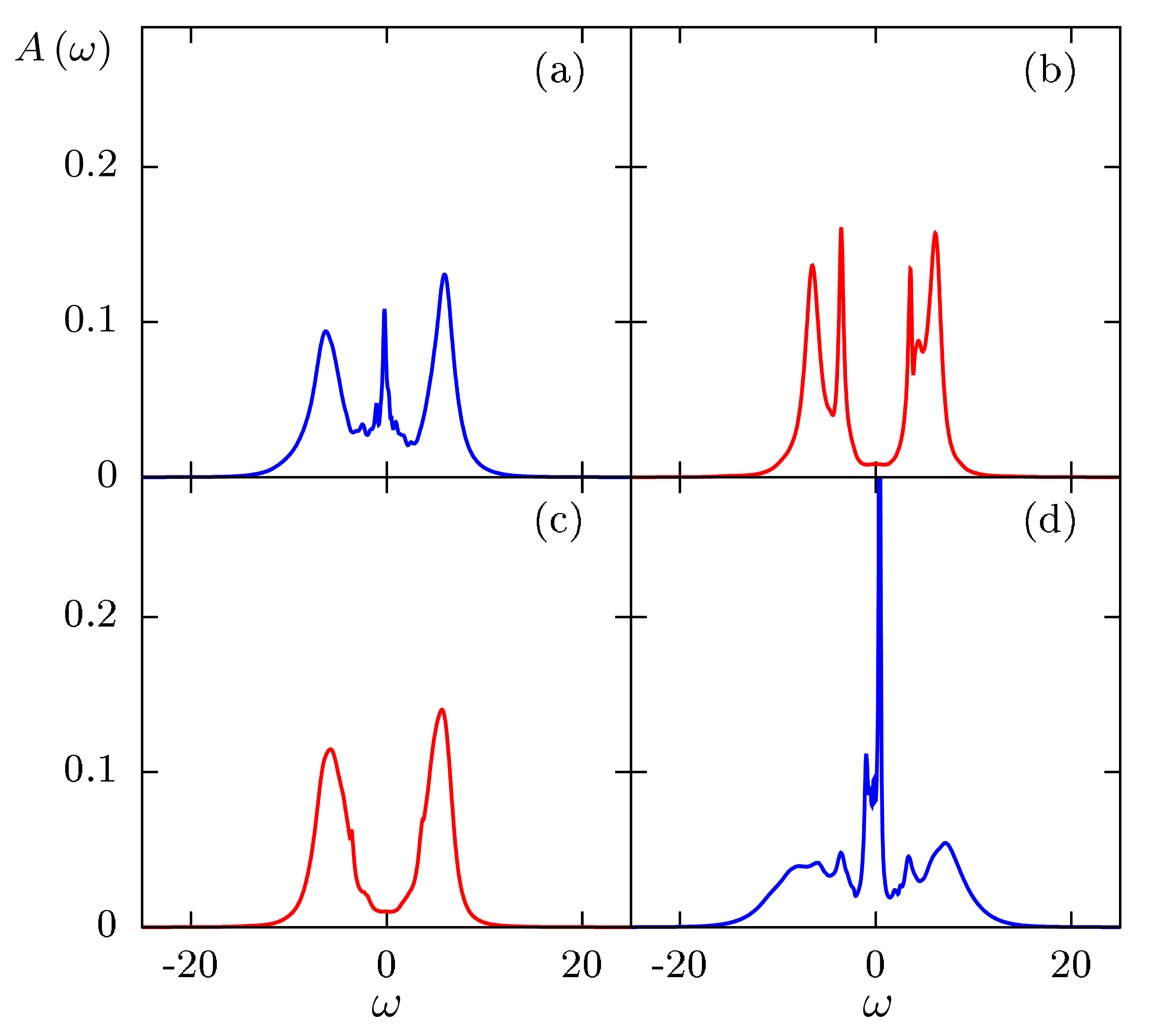}
\caption{Local spectral functions $A_i(\omega, t)$ on four different atoms at a given instant of the DMFT-MD simulations. The Hubbard parameter is $U = 2.29 W_0$.}
\label{fig:local-Aw2}
\end{figure}

\begin{figure}[t]
\includegraphics[width=0.9\linewidth]{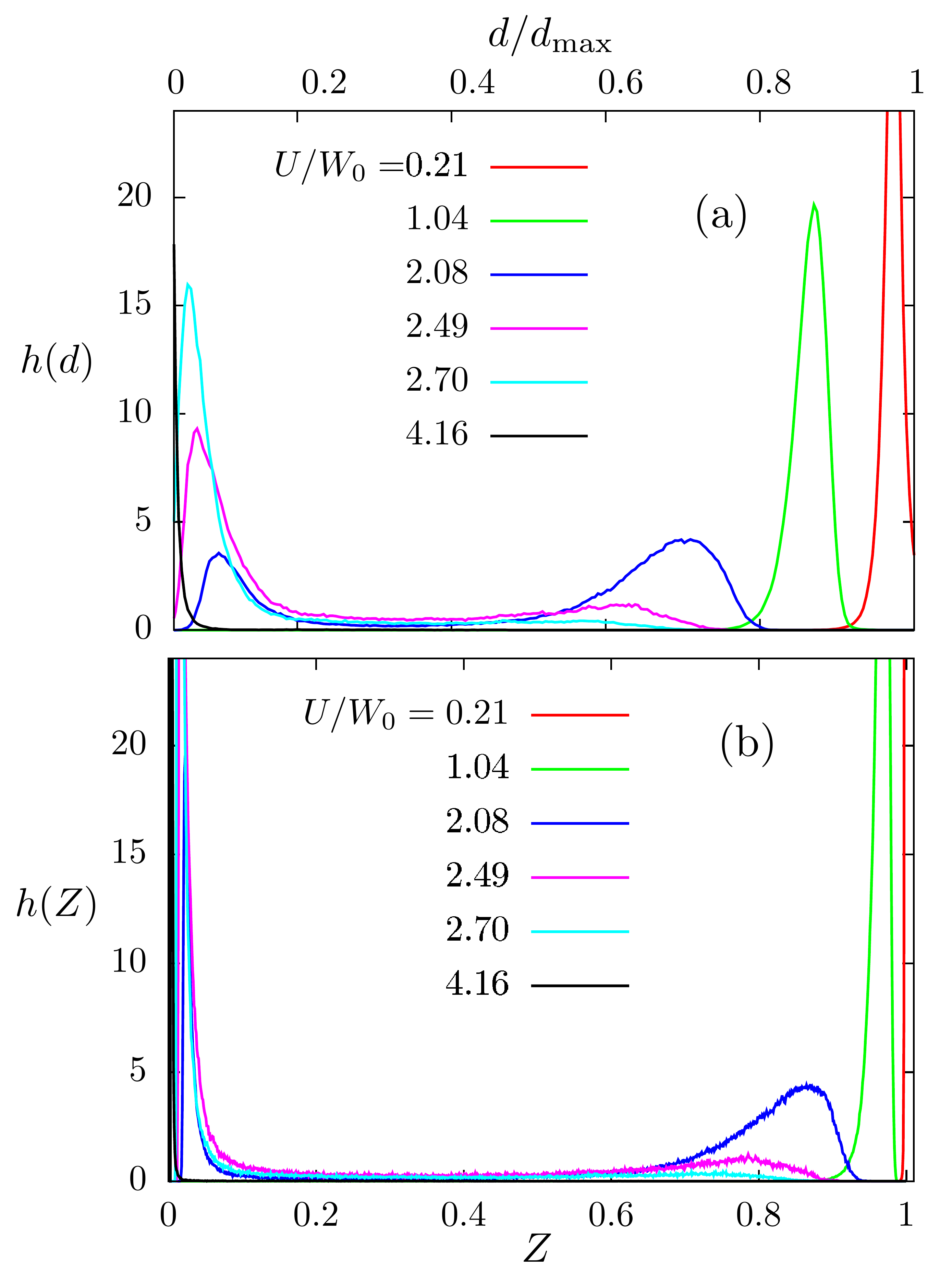}
\caption{Histogram of (a) local double occupancy $d_i$ and (b)~quasi-particle weight $Z_i$ at varying Hubbard repulsion obtained from the DMFT-MD simulations.  The Mott transition occurs at a critical $U_c \sim 3.1 W_0$. The distribution function $h(d)$ in the intermediate $U$ regime exhibits a characteristic bimodal structure.}
\label{fig:hist2}
\end{figure}

Electronically, the heterogeneity of the intermediate metallic states manifests itself in the rather broad optical absorption spectrum discussed above. Indeed, detailed analysis finds two distinct types of local spectral functions $A_i(\omega; t)$, one with the quasi-particle peak at $\omega = 0$ and the other without, coexisting in this intermediate correlated regime; see Fig.~\ref{fig:local-Aw2}. This atomic-scale heterogeneity is also demonstrated from the histograms of the local double-occupancy $d_i$ and quasi-particle weight $Z_i$ shown in Fig.~\ref{fig:hist2}. Take the double-occupancy as example, the distribution function $h(d)$ in the insulating dimer phase and the Mott phase is marked by a single pronounced peak at $d \sim d_{\rm max} = 0.25$ (for half-filling) and $d \sim 0$, respectively. On the other hand, the distribution function exhibits a bimodal structure with two well separated peaks in the intermediate-$U$ regime, e.g. the case of $U = 2.08 W_0$. This bimodal distribution is indicative of two species of atoms, one associated with a persistent quasi-dimers and the other with quasi-localized electrons. 

The finite dc conductivity of the intermediate state can also be understood as originating from quasi-dimers of variable bond-lengths in an extended clusters. It is also worth noting that the coexistence of two types of atoms also highlight the first-order nature of the Mott transition. Yet, the dynamical nature due to the mobility of atoms in a liquid state results in a smooth crossover, instead of a sharp discontinuity typical of a first-order transition. The frequent atomic collisions transform the atoms from one type to the other. The resultant intermediate metallic state is thus more of a homogeneous mixture of atoms of different species, instead of an inhomogeneous collection of different phases as in a phase-separated state. 

\begin{figure}
\includegraphics[width=0.95\linewidth]{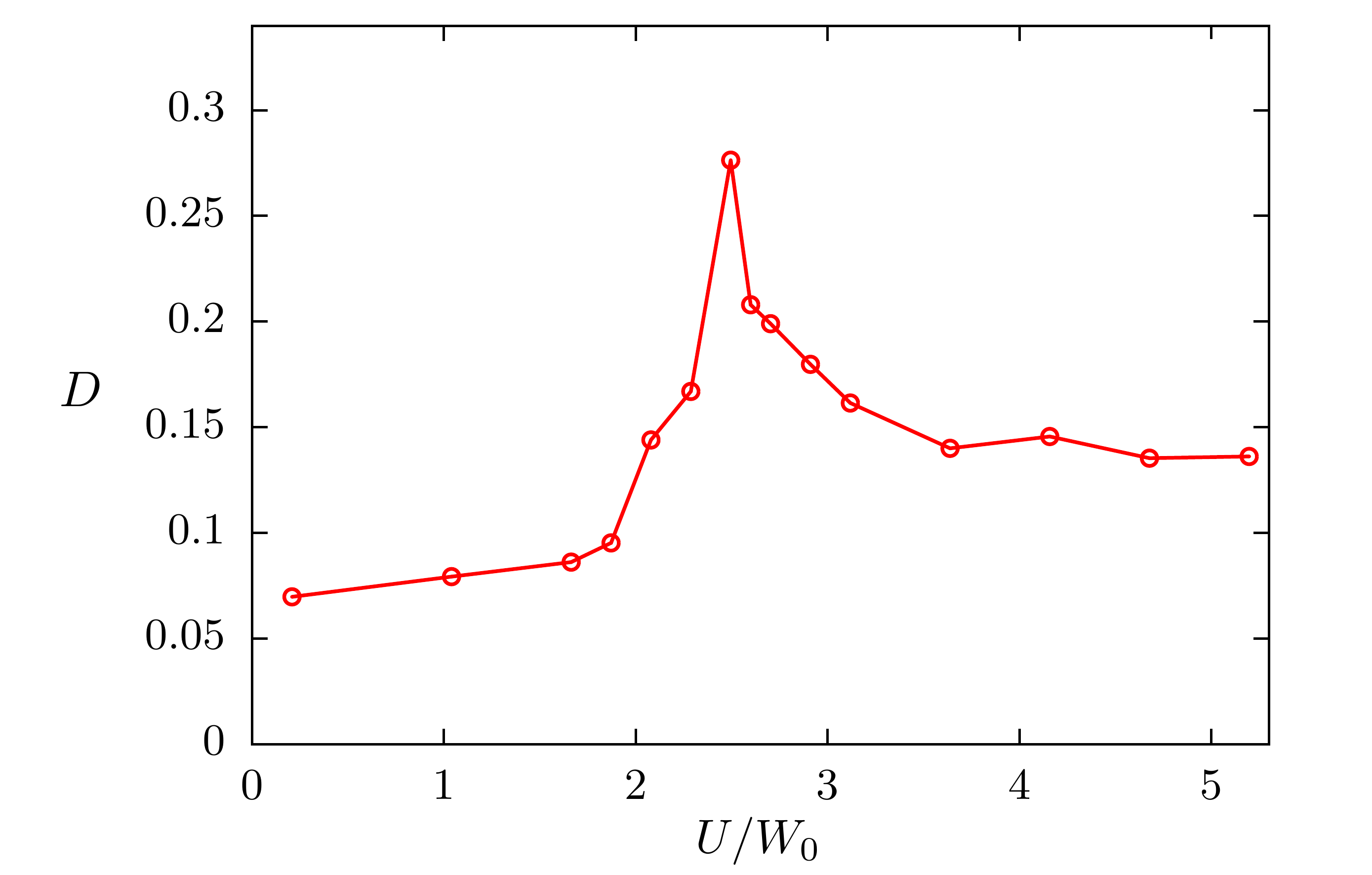}
\caption{The self-diffusion coefficient $D$, computed using Eq.~(\ref{eq:diffusion}), is plotted as a function the Hubbard parameter $U$. }
\label{fig:diffusion}
\end{figure}

Finally, as the interatomic forces depend strongly on the nature of the chemical bonding, the above three distinct regimes are expected to exhibit different atomic transport behaviors. To this end, we use Eq.~(\ref{eq:diffusion}) to compute the self-diffusion coefficient $D$ from the DMFT-MD simulations and plot it as a function of $U$ in Fig.~\ref{fig:diffusion}. The diffusion coefficient remains approximately constant for both the small-$U$ dimer phase and the large-$U$ Mott phase. This result is expected as there is no significant changes in the nature of inter-molecular or interatomic forces, respectively, in these two regimes. The slight increase of $D$ in the dimer phase could be attributed to the gradual dissociation of dimers. 

Most notably, the diffusion coefficient exhibits a pronounced maximum in the intermediate $U$ regime that corresponds to the heterogeneous metallic state. A general theory of the self-diffusion coefficient of a correlated atomic liquid has recently been developed in Ref.~\cite{cheng22}. In particular, starting from the Mott insulating phase where the interatomic interaction is purely repulsive, the enhancement of the atomic diffusivity is attributed to a reduced radius of the effective repulsive core when $U$ is decreased. Importantly, this reduction is caused by the attractive interatomic interaction due to delocalized electrons. Once the attractive cohesion dominants the interatomic forces, the diffusion coefficient then becomes smaller as $U$ is further reduced. The subtle interplay of the attractive cohesive force and the short-range repulsion thus leads to a maximum in the diffusion coefficient in the vicinity of a Mott transition. 
The above scenario also applies to our case. Specifically, the diffusion maximum here indicates that the diffusivity is enhanced for atoms moving in the metallic clusters originating from the partially dissociated dimers. The situation is more complicated here because of the formation of dimers


\section{Discussion}

\label{sec:discussion}

To summarize, we have presented a fundamental theoretical picture of Mott metal-insulator transitions in simple fluid metals based on quantum MD simulations in which the electronic structure is solved by a real-space formulation of the dynamical mean-field theory (DMFT). Contrary to quantum MD approaches that rely on a self-consistent single-electron theory such as the DFT or Gutzwiller approximation, the DMFT-MD method not only can account for essential electron correlation effects such as Mott transition, but also provide valuable information on incoherent electronic excitations such as the formation of Hubbard bands. The single-band Hubbard model is generalized to an atomic liquid system within the tight-binding molecular dynamics (TBMD) formalism. Moreover, this atomic liquid system characterized by a correlated $s$-band, dubbed the Hubbard liquid model, also serves as a minimum theoretical model for studying metal-insulator transition in the alkali fluids. 

Applying the DMFT-MD simulations to this model liquid, we uncover two different scenarios of liquid-state Mott transitions, which are distinguished by whether localized atomic structures such as molecules are stabilized by the chemical bonding in the electronically delocalized phase. In the absence of such local structures, which in our model corresponds to an ultra-short repulsive core, the electron-mediated cohesive forces give rise to the formation of extended atomic clusters which resemble the atomic structures of amorphous solids. Increasing electron correlation leads to localization of electrons and the breakup of the atomic clusters. The example, we believe, offers the fundamental picture of a liquid-state Mott transition which is to be contrasted with those in crystalline or amorphous systems. Electronically, the Mott transition here is characterized by the transfer of spectral weight from the quasi-particle peak to high-energy Hubbard bands, similar to the standard DMFT scenarios. Despite the similarities, our simulations uncover an intriguing heterogeneous state in the strongly correlated metallic regime, which is a dynamical mixture of metallic clusters diluted by atoms of localized electrons.

On the other hand, for a repulsive potential with a range comparable to that of electron hopping, the strong repulsion presents the formation of large clusters. The resultant chemical bonding, which of the valence bond nature, favors the formation of diatomic molecules in the ``conducting" phase. The spectrum of the resultant molecular liquid is characterized by a spectral gap separating the bonding from the anti-bonding states. Enhanced electron correlation results in the dissociation of molecules and the emergence of an intermediate metallic phase. The electronic spectral function of this intermediate state is characterized by a quasi-particle peak at the Fermi liquid. Upon further increasing the electron correlation, the intermediate state undergoes the Mott transition with the development of a three-peak structure in the spectral function. More generally, this scenario of Mott transition highlights the nontrivial interplay of local atomic structures stabilized by valence bonding and the correlation-induced localization of electrons. 

Interestingly, the dissociation of diatomic molecules has been observed in DFT-MD simulations of hydrogen liquids at high temperatures. Despite being the simplest element of the periodic table, hydrogen continues to fascinate researchers in condensed-matter physics, energy application, and planetary science~\cite{ashcroft00,mcmahon12}. This is because hydrogen exhibits a rich phase diagram~\cite{ackland15}, which contains at least four different solid phases as well as several liquid phases. Furthermore, it is predicted to display remarkable properties such as low-temperature quantum fluidity and high-pressure superconductivity~\cite{bonev04,richardson97,babaev04,cudazzo08}.

In particular, it was conjectured by Wigner and Huntington in the early days of quantum mechanics that hydrogens might undergo a { liquid-liquid insulator to metal transition} with increasing pressure~\cite{wigner35}. The liquid metallic hydrogen at high pressures is also special in the sense that it is an atomic liquid. Indeed, all other phases of hydrogen, including the solid phases and the insulating liquid, consist of hydrogen H$_{\rm 2}$ molecule as the basic unit. The insulator-to-metal transition thus also corresponds to the dissociation of the hydrogen molecules, which results in the diminishing of the molecule peak in the distribution function, a phenomenon already observed in both GA-MD~\cite{chern17} as well as DMFT-MD simulations of the liquid Hubbard model discussed above. Further works by combing the DMFT-MD methods with {\em ab initio} modeling  are required to understand the role of electron correlation in the liquid-liquid transition of hydrogens. 



\begin{figure}[t]
\includegraphics[width=1.0\columnwidth]{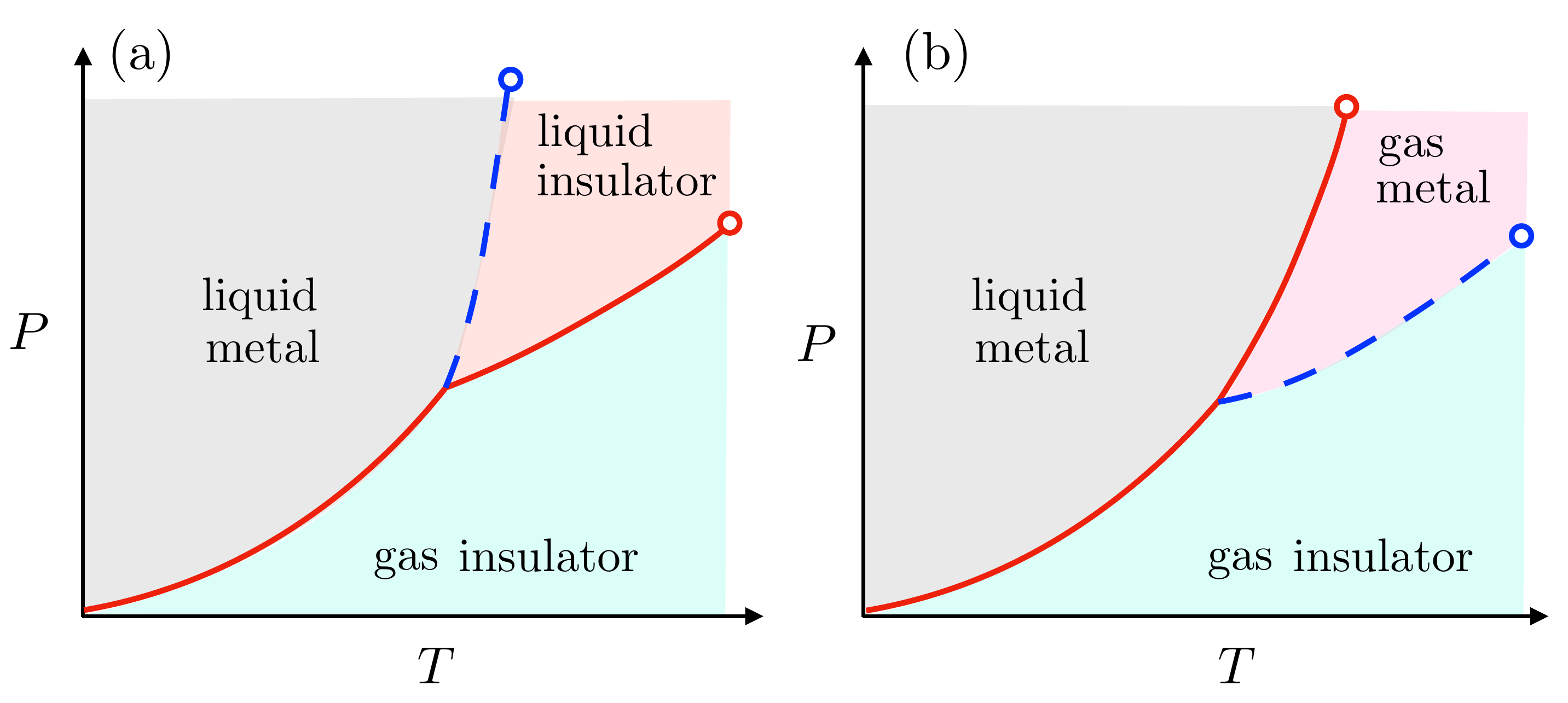}
\caption{Phase diagrams for liquid metals as proposed by Zel'dovich and Landau~\cite{zeldovich44}. In both cases, the first-order MIT (blue dashed line) is separated from the first-order liquid gas transition (red solid line). In panel~(a), the MIT occurs in the liquid phase, while it is in the gas phase in scenario~(b). The two circles indicate the corresponding critical end points. 
}
\label{fig:phase}  
\end{figure}

The Mott transition that involves extended conducting clusters is also relevant to the metal-insulator transition in expanding alkali fluids such as liquid cesium or rubidium. In addition to its fundamental importance, part of the interest is motivated by the fact that liquid metals are attractive as thermodynamic working fluids for many advanced energy technologies because they have high latent heats of vaporization and high heat transfer coefficients. In typical experiments, the MIT is driven by reducing the atomic density through the expansion of the metallic fluid.  An intriguing scenario, proposed long ago by Zel'dovich and Landau~\cite{zeldovich44}, posits that the first-order MIT is separated from the first-order liquid-gas transition, although the MIT line could be on either the liquid or the vapor side of the phase diagram; see Fig.~\ref{fig:phase}.  While this scenario seems to be confirmed in liquid metals such as mercury and iron, the MIT in alkali liquids is found to occur at densities close to the liquid-gas critical point. Since the transitions occur at rather high temperatures and pressure ($T_c \approx 1924$ K for Cs and 2017 K for Rb), so far experiments cannot unambiguously distinguish the two transitions due to the extreme conditions.

Contrary to polyvalent liquid metal systems, where the MIT is likely due to the combination of band splitting (the Bloch-Wilson mechanism) and Anderson localization, electron correlation of the Mott-Hubbard type is believed to play a crucial part in the density-driven metal-non-metal transition in alkali liquids. Indeed, with one electron per atom in the $s$-band, the alkali metal is the ideal platform to realize the Hubbard model. Experimental evidence of strong electron correlation comes from various measurements of magnetic properties in the vicinity of the critical point. By measuring the magnetic susceptibility $\chi_m$ along the liquid-vapor coexistence curve in an expanded cesium, Freyland observed an increasing enhancement of $\chi_m$ starting at a density that is three times the critical density $\rho_c$ in the liquid phase~\cite{freyland79}. The increase of the magnetic susceptibility saturates at the Curie value at a density $\sim 2 \rho_c$. As pointed out in several ensuing works, the enhancement of $\chi_m$ and its limitation by the Curie-law strongly are best explained by a renormalized $s$-band caused by electron correlation~\cite{rose81,chapman88,redmer93}. 

The enhancement of effective electron mass as the band narrows was also supported by NMR Knight shift and optical reflectivity measurements of expanded liquid cesium~\cite{warren89,knuth97}. Moreover, the Korringa ratio obtained from dynamical susceptibility measurement reveals an incipient antiferromagnetic spin-correlation in the liquid cesium metal, which is another signature of electron correlation~\cite{warren89,warren93}. These experiments had motivated theoretical investigations on the lattice-gas models for liquid metals that explicitly include electron correlation. In the so-called lattice-gas Hubbard model, the ions in this model are modeled by classical gas particles on a lattice, and electrons are described by the associated TB-Hubbard model on connected ions. Both mean-field and Monte Carlo methods were used to sample the ionic configurations, while the electron problem was solved based on mean-field type methods, including the Gutzwiller approximation. While several predictions of these approaches agreed with experiments on alkali liquids, the  introduction of an underlying lattice results in artifacts which might over-emphasize the percolation mechanism. Dynamical properties such as atomic diffusion, and how they are affected by the electrons, are also beyond the Monte Carlo approaches. 

On the other hand, both classical and DFT MD methods have been applied to study the MIT in alkali liquids, although, as discussed in Sec.~\ref{sec:intro}, the crucial electron correlation effect is not properly accounted for in these approaches. The static atomic structure and self-diffusion coefficients obtained from these MD studies agreed reasonably with the experiments. One important structural feature that is successfully captured by the MD simulations is that the intensity of the first peak in structure factor $S(Q_1)$ is reduced with decreasing density, yet the corresponding wave vector $Q_1$ remains roughly the same. These results clearly indicate that, instead of a uniform expansion, the reduced atomic density results in a less packed structure with atoms forming clusters that are loosely connected to each other. This observation is basically consistent with the Mott transition scenario obtained in our DMFT-MD simulations. Despite the highly disordered atomic configuration in an expanding alkali fluid, the persistence of such metallic clusters implies the crucial role of electron correlation in the MIT of these liquid systems.

\bigskip

\begin{acknowledgments}
We thank C.~D.~Batista, V. Dobrosavljevi\'c, and G.~Kotliar for useful discussions. We also thank Chen~Cheng for collaboration on related projects and numerous insightful comments on the interplay between atom dynamics and Mott transitions in the liquid state. This work is partially supported by the Center for Materials Theory as a part of the Computational Materials Science (CMS) program, funded by the US Department of Energy, Office of Science, Basic Energy Sciences, Materials Sciences and Engineering Division. The authors also acknowledge the support of Advanced Research Computing Services at the University of Virginia.
\end{acknowledgments}


\begin{thebibliography}{99}


\bibitem{mott90} N. F. Mott, {\em Metal-Insulator Transitions} (Taylor \& Francis, New York, 1997).

\bibitem{imada98} M. Imada, A. Fujimori, and Y. Tokura, Metal-insulator transitions, Rev. Mod. Phys. {\bf 70}, 1039 (1998).

\bibitem{dobrosavljevic12} V. Dobrosavljevi\'c, {\em Introduction to Metal-Insulator Transitions} (Oxford University Press, Oxford, 2012). 


\bibitem{anderson58} P. W. Anderson, Absence of Diffusion in Certain Random Lattices, Phys. Rev. {\bf 109}, 1492 (1958).

\bibitem{lee85} P. A. Lee and T. V. Ramakrishnan, Disordered electronic systems, Rev. Mod. Phys. {\bf 57}, 287 (1985).


\bibitem{peierls37} N. Mott and R. Peierls, Discussion of the paper by de Boer and Verwey, Proc. R. Soc. , Ser.  {\bf A49}, 72 (1937).

\bibitem{mott49} N. F. Mott, The Basis of the Electron Theory of Metals, with Special Reference to the Transition Metals, Proc. Phys. Soc. London Ser. {\bf A62}, 416 (1949).


\bibitem{paalanen88} M. A. Paalanen, J. E. Graebner, R. N. Bhatt, and S. Sachdev, Thermodynamic Behavior near a Metal-Insulator Transition, Phys. Rev. Lett. {\bf 61}, 597 (1988).

\bibitem{milovanovic89} Milica Milovanovi\'c, Subir Sachdev, and R. N. Bhatt, Effective-field theory of local-moment formation in disordered metals, Phys. Rev. Lett. {\bf 63}, 82 (1989).

\bibitem{bhatt92} R. N. Bhatt and D. S. Fisher, Absence of Spin Diffusion in Most Random Lattices, Phys. Rev. Lett. {\bf 68}, 3072 (1992).

\bibitem{belitz94} D. Belitz and T. R. Kirkpatrick, The Anderson-Mott transition, Rev. Mod. Phys. {\bf 66}, 261 (1994).

\bibitem{dobrosavljevic97} V. Dobrosavljevi\'c and G. Kotliar, Mean Field Theory of the Mott-Anderson Transition, Phys. Rev. Lett. {\bf 78}, 3943 (1997).

\bibitem{miranda01} E. Miranda and V. Dobrosavljevi\'c, Localization-Induced Griffiths Phase of Disordered Anderson Lattices, Phys. Rev. Lett. {\bf 86}, 264 (2001).

\bibitem{miranda05} E. Miranda and V. Dobrosavljevi\'c, Disorder-driven non-Fermi liquid behaviour of correlated electrons, Rep. Prog. Phys. {\bf 68}, 2337 (2005).

\bibitem{yamamoto20} R. Yamamoto, T. Furukawa, K. Miyagawa, T. Sasaki, K. Kanoda, and T. Itou, Electronic Griffiths Phase in Disordered Mott-Transition Systems, Phys. Rev. Lett. {\bf 124}, 046404 (2020).


\bibitem{kikoin66} I. K. Kikoin, A. P. Senchenkov, E. V. Gel'man, M. M. Korsunskii, S. P. Naurzakov, Electrical conductivity and density of a metal vapor, Sov. Phys. JETP, {\bf 22}, 89 (1966).

\bibitem{hensel68} F. Hensel, E. U. Franck, Metal-nonmetal transition in dense mercury vapor, Rev. Mod. Phys. {\bf 40},  697 (1968).

\bibitem{yonezawa82} F. Yonezawa and T. Ogawa, Metal-Nonmetal Transitions in Expanded Fluids, Suppl. Prog. Theo. Phys. {\bf 72}, 1 (1982).

\bibitem{hensel89} F. Hensel and H. Uchtmann, The Metal-Insulator Transition in Expanded Fluid Metals, Annu. Rev. Phys. Chem. {\bf 40}, 61 (1989).

\bibitem{mott66} N. F. Mott, The electrical properties of liquid mercury, Philos. Mag. {\bf 13},  989 (1966).

\bibitem{ziman61} J. M. Ziman, A theory of the electrical properties of liquid metals. I: The monovalent metals, {\bf 6}, 1013 (1961).

\bibitem{cohen74} M. H. Cohen, J. Jortner, Conduction regimes in expanded liquid mercury, Phys. Rev. A {\bf 10},  978  (1974).






\bibitem{kirkpatrick73} S. Kirkpatrick, Percolation and Conduction, Rev. Mod. Phys. {\bf 45}, 574 (1973).

\bibitem{odagaki75} T. Odagaki, N. Ogita, and H. Matsuda, Percolation approach to the metal-insulator transition in super-critical fluid metals, J. Phys. Soc. Jpn. {\bf 39}, 618 (1975).

\bibitem{phelps76} D. J. Phelps and C. P. Flynn, Metal-insulator transition in rare-gas-alkali-metal thin films, Phys. Rev. B {\bf 14}, 5279 (1976).

\bibitem{franz84} J. R. Franz, Metal-insulator transition in expanded alkali-metal fluids and alkali-metal -- rare-gas films, Phys. Rev. B {\bf 29}, 1565 (1984).


\bibitem{nield91} V. M. Nield, M. A. Howe, and R. L. McGreevy, The metal-non-metal transition in expanded caesium, J. Phys.: Condens. Matter {\bf 3}, 7519 (1991).

\bibitem{tarazona95} P. Tarazona, E. Chacon, and J. P. Hernandez, Simple Model for the Phase Coexistence and Electrical Conductivity of Alkali Fluids, Phys. Rev. Lett. {\bf 74}, 142 (1995). 

\bibitem{arai99} T. Arai and R. L. McGreevy, Percolation Aspects of the Metal-Insulator Transition in Expanded Cs, Phys. Chem. Liquids, {\bf 34}, 455 (1999).


\bibitem{cadien13} A. Cadien, Q. Y. Hu, Y. Meng, Y. Q. Cheng, M. W. Chen, J. F. Shu, H. K. Mao, and H. W. Sheng, First-Order Liquid-Liquid Phase Transition in Cerium, Phys. Rev. Lett. {\bf 110}, 125503 (2013).

%

\bibitem{yonezawa73} F. Yonezawa, M. Watabe, Electron correlation and metal-nonmetal transition in a disordered binary system, Phys. Rev. B {\bf 8}, 4540 (1973).

\bibitem{yonezawa74} F. Yonezawa, M. Watabe, M. Nakamura, Y. Ishida, Electron correlations and metal-insulator transition in structurally disordered systems: General theory and application to supercritical alkali metals, Phys. Rev. B {\bf 10}, 2322 (1974).


\bibitem{jungst85} S. J\"ungst, B. Knuth, and F. Hensel, Observation of Singular Diameters in the Coexistence Curves of Metals, Phys. Rev. Lett. {\bf 55}, 2160 (1985). 

\bibitem{winter87} R. Winter, T. Bodensteiner, W. Gl\"aser, and F. Hensel, The static structure factor of cesium over the whole liquid range up to the critical point, Ber. Bunsenges. Phys. Chem. {\bf 91}, 1327 (1987).

\bibitem{hensel89b} F. Hensel, Experiments on Expanded Metals in the Metal-Nonmetal Transition Region, Phys. Scr. {\bf T25}, 283 (1989).

\bibitem{hensel85} F. Hensel, S. Jungst, F. Noll, and R. Winter, in {\em Localisation and Metal Insulator Transitions}, edited by D. Adler and H. Fritsche, 53. (Plenum Press, 1985).


\bibitem{freyland79} W. Freyland, Magnetic susceptibility of metallic and nonmetallic expanded fluid cesium, Phys. Rev. B {\bf 20}, 5104 (1979).

\bibitem{hanany83} U. El-Hanany, G. F. Brennert, and W. W. Warren, Jr., Enhanced Paramagnetism and Spin Fluctuations in Expanded Liquid Cesium, Phys. Rev. Lett. {\bf 50}, 540 (1983).


%

\bibitem{binder04} K. Binder, J. Horbach, W. Kob, W. Paul, and F. Varnik, Molecular dynamics simulations, J. Phys.: Condens. Matter {\bf16}, S429 (2004).

\bibitem{allen89} M. Allen and D. Tildesley, {\em Computer simulation of liquids} (Oxford University Press, New York, 1989).

\bibitem{rapaport04} D. C. Rapaport, {\em The Art of Molecular Dynamics Simulation} (Cambridge University Press, New York, 2004).

\bibitem{fermi55} E. Fermi, J. Pasta, S. Ulam, Studies of Nonlinear Problems, Document LA-1940. Los Alamos National Laboratory (1955).

\bibitem{mackerell04} A. D. Mackerell, Empirical force fields for biological macromolecules: Overview and issues, J. Comput. Chem. {\bf 25} 1584 (2004).



\bibitem{car85} R. Car and M. Parrinello, Unified Approach for Molecular Dynamics and Density-Functional Theory, Phys. Rev. Lett. {\bf 55}, 2471 (1985).

\bibitem{payne92} M. C. Payne, M. P. Teter, D. C. Allan, T. A. Arias, and J. D. Joannopoulos, Iterative minimization techniques for {\em ab initio} total-energy calculations: molecular dynamics and conjugate gradients, Rev. Mod. Phys. {\bf 64}, 1045 (1992).

\bibitem{kresse93} G. Kresse and J. Hafner, {\em Ab initio} molecular dynamics for liquid metals, Phys. Rev. B {\bf 47}, 558(R) (1993).

\bibitem{tuckerman02} M. Tuckerman,  Ab initio molecular dynamics: basic concepts, current trends and novel applications, J. Phys.: Cond. Matter {\bf 14}, R1297 (2002).


\bibitem{attaccalite08} C. Attaccalite and S. Sorella, { Stable Liquid Hydrogen at High Pressure by a Novel Ab Initio Molecular-Dynamics Calculation}, Phys. Rev. Lett. {\bf 100}, 114501 (2008).


\bibitem{marx09} D. Marx and J. Hutter, {\em Ab initio molecular dynamics: basic theory and advanced methods} (Cambridge University Press, Cambridge, 2009).





\bibitem{gordon88} M. Head-Gordon and J. A. Pople, Optimization of wave function and geometry in the finite basis Hartree-Fock method, J. Phys. Chem. {\bf 92}, 3063 (1988).

\bibitem{field91} M. J. Field, Constrained optimization of ab initio and semiempirical Hartree-Fock wave functions using direct minimization or simulated annealing, J. Phys. Chem. {\bf 95}, 5104 (1991).




\bibitem{hohenberg64} P. Hohenberg,  and W. Kohn, Inhomogeneous Electron Gas, Phys. Rev. {\bf 136}, B864 (1964).

\bibitem{kohn65} W. Kohn and L. J. Sham, Self-Consistent Equations Including Exchange and Correlation Effects, Phys. Rev. {\bf 140}, A1133 (1965).


\bibitem{jones15} R. O. Jones, Density functional theory: Its origins, rise to prominence, and future, Rev. Mod. Phys. {\bf 87}, 897 (2015).


\bibitem{perdew91} J. P. Perdew and Y. Wang, Accurate and simple analytic representation of the electron-gas correlation energy, Phys. Rev. B {\bf 45}, 13244 (1992).

\bibitem{becke88} A. D. Becke, Density-functional exchange-energy approximation with correct asymptotic behavior, Phys. Rev. A {\bf 38}, 3098 (1988).

\bibitem{perdew96} J. P. Perdew, K. Burke, and M. Ernzerhof, Generalized Gradient Approximation Made Simple, Phys. Rev. Lett. {\bf 77}, 3865 (1996).


\bibitem{perdew81} J. P. Perdew and A. Zunger, Self-interaction correction to density-functional approximations for many-electron systems, Phys. Rev. B {\bf 23}, 5048 (1981).

\bibitem{tsuneda14} T. Tsuneda and K. Hirao, Self-interaction corrections in density functional theory, J. Chem. Phys. {\bf 140}, 18A513 (2014).


\bibitem{anisimov91} V. I. Anisimov and O. Gunnarsson, Density-functional calculation of effective Coulomb interactions in metals, Phys. Rev. B {\bf 43}, 7570 (1991).

\bibitem{anisimov91b} V. I. Anisimov, J. Zaanen, and O. K. Andersen, Band theory and Mott insulators: Hubbard $U$ instead of Stoner $I$, Phys. Rev. B {\bf 44}, 943  (1991).




\bibitem{fazekas99} P. Fazekas, {\em Lecture Notes on Electron Correlation and Magnetism} (World Scientific, Singapore, 1999)

\bibitem{hewson93} A. C. Hewson, {\em The Kondo Problem to Heavy Fermions} (Cambridge University Press, New York, 1993).





\bibitem{gutzwiller63} M.~C.~Gutzwiller, Effect of correlation on the ferromagnetism of transition metals, Phys. Rev. Lett. {\bf 10}, 159 (1963).

\bibitem{gutzwiller64} M.~C.~Gutzwiller, Effect of Correlation on the Ferromagnetism of Transition Metals, Phys. Rev. {\bf 134} A923 (1964).

\bibitem{gutzwiller65} M.~C.~Gutzwiller, Correlation of Electrons in a Narrow $s$~Band, Phys. Rev. {\bf 137} A1726 (1965).




\bibitem{brinkman70} W. F. Brinkman and T. M. Rice, Application of Gutzwiller's variational method to the metal-insulator transition, Phys. Rev. B {\bf 2}, 4302 (1970).

\bibitem{vollhardt84} D. Vollhardt, Normal $^3$He: an almost localized Fermi liquid, Rev. Mod. Phys. {\bf 56}, 99 (1984).



\bibitem{chern17} G.-W. Chern, K. Barros, C. D. Batista, J. Kress, and G. Kotliar, Mott transition in a metallic liquid -- Gutzwiller molecular dynamics simulations, Phys. Rev. Lett. {\bf 118}, 226401 (2017).

\bibitem{suwa19} H. Suwa, J. S. Smith, N. Lubbers, C. D. Batista, G.-W. Chern, and K. Barros, Machine learning for molecular dynamics with strongly correlated electrons, Phys. Rev. B {\bf 99}, 161107(R) (2019).

\bibitem{chern19} G.-W. Chern, Kinetics of thermal Mott transitions in the Hubbard model, arXiv:1907.05880 (2019).




\bibitem{kotliar86} G. Kotliar and A. E. Ruckenstein, New functional integral approach to strongly correlated Fermi systems: The Gutzwiller approximation as a saddle point, Phys. Rev. Lett. {\bf 57}, 1362 (1986).

\bibitem{bunemann07} J. B\"{u}nemann and F. Gebhard, Equivalence of Gutzwiller and Slave-Boson Mean-Field Theories for Multiband Hubbard Models, Phys. Rev. B {\bf 76}, 193104 (2007).

\bibitem{lechermann07} F. Lechermann, A. Georges, G. Kotliar, and O. Parcollet, Rotationally Invariant Slave-Boson Formalism and Momentum Dependence of the Quasiparticle Weight, Phys. Rev. B {\bf 76}, 155102 (2007).


\bibitem{lanata17} N. Lanat\'{a}, Y. Yao, X. Deng, V. Dobrosavljevi\'{c}, and G. Kotliar, Slave Boson theory of orbital differentiation with crystal field effects: Application to UO$_2$, Phys. Rev. Lett. {\bf 118}, 126401 (2017).



\bibitem{kotliar04} G. Kotliar and D. Vollhardt, Strongly Correlated Materials: Insights From Dynamical Mean-Field Theory, Phys. Today {\bf 57}, 53 (2004).

\bibitem{georges96} A. Georges, G. Kotliar, W. Krauth, and M. J. Rozenberg, Dynamical mean-field theory of strongly correlated fermion systems and the limit of infinite dimensions, Rev. Mod. Phys. {\bf 68}, 13 (1996).

\bibitem{kotliar06} G. Kotliar, S. Y. Savrasov, K. Haule, V. S. Oudovenko, O. Parcollet, and C. A. Marianetti, Electronic structure calculations with dynamical mean-field theory, Rev. Mod. Phys. {\bf 78}, 865 (2006).





\bibitem{metzner87} W. Metzner and D. Vollhardt, Ground-state properties of correlated fermions: Exact analytic results for the Gutzwiller wave function, Phys. Rev. Lett. {\bf 59}, 121 (1987).

\bibitem{metzner89} W. Metzner and D. Vollhardt, Correlated Lattice Fermions in $d=\infty$ Dimensions, Phys. Rev. Lett. {\bf 62}, 324 (1989).

\bibitem{hartmann89} E. M\"uller-Hartmann, The Hubbard model at high dimensions: some exact results and weak coupling theory, Z. Phys. B {\bf 76}, 211 (1989).





\bibitem{potthoff99} M. Potthoff and W. Nolting, Surface metal-insulator transition in the Hubbard model, Phys. Rev. B {\bf 59}, 2549 (1999).

\bibitem{ishida09} H. Ishida and A. Liebsch, Embedding approach for dynamical mean-field theory of strongly correlated heterostructures, Phys. Rev. B {\bf 79}, 045130 (2009).

\bibitem{freericks04} J. K. Freericks, Dynamical mean-field theory for strongly correlated inhomogeneous multilayered nanostructures, Phys. Rev. B {\bf 70}, 195342 (2004).

\bibitem{tran07} M.-T. Tran, Statistics of local density of states in the Falicov-Kimball model with local disorder, Phys. Rev. B {\bf 76}, 245122 (2007). 

\bibitem{okamoto08} Satoshi Okamoto, Nonlinear Transport through Strongly Correlated Two-Terminal Heterostructures: A Dynamical Mean-Field Approach, Phys. Rev. Lett. {\bf 101}, 116807 (2008).

\bibitem{helmes08} R. W. Helmes, T. A. Costi, and A. Rosch, Mott transition of Fermionic atoms in a three-dimensional optical trap, Phys. Rev. Lett. {\bf 100}, 056403 (2008).

\bibitem{snoek08} M. Snoek, I. Titvinidze, C. T\"oke, K. Byczuk, and W, Hofstetter, Antiferromagnetic order of strongly interacting fermions in a trap: real-space dynamical mean-field analysis, New J. Phys. {\bf 10}, 093008 (2008).


\bibitem{gorelik10} E. V. Gorelik, I. Titvinidze, W. Hofstetter, M. Snoek, and N. Bl\"umer, Neel transition of lattice fermions in a harmonic trap: A real-space dynamic mean-field study, Phys. Rev. Lett. {\bf 105}, 065301(2010).



\bibitem{hubbard63} J. Hubbard, Electron correlations in narrow energy bands, Proc. R. Soc. A {\bf 276}, 238 (1963).

\bibitem{hubbard64} J. Hubbard, Electron correlations in narrow energy bands III. An improved solution, Proc. R. Soc. A {\bf 281}, 401 (1964).

\bibitem{kanamori63} J. Kanamori, Electron Correlation and Ferromagnetism of Transition Metals, Prog. Theor. Phys. {\bf 30}, 275 (1963).


\bibitem{tasaki98} H. Tasaki, The Hubbard model -- an introduction and selected rigorous results, J. Phys.: Condens. Matter {\bf 10}, 4353 (1998).


\bibitem{arovas21} D. P. Arovas, E. Berg, S. Kivelson, S. Raghu, The Hubbard Model, arXiv:2103.12097, To appear in Annu. Rev. Condens. Matter Phys. (2021).



\bibitem{leblanc15} J. P. F. LeBlanc, A. E. Antipov, F. Becca, I. W. Bulik, G. Kin-Lic Chan, C.-M. Chung, Y. Deng, M. Ferrero, T. M. Henderson, C. A. Jimenez-Hoyos, E. Kozik, X.-W. Liu, A. J. Millis, N. V. Prokof'ev, M. Qin, G. E. Scuseria, H. Shi, B. V. Svistunov, L. F. Tocchio, I.?S. Tupitsyn, S. R. White, S. Zhang, B.-X. Zheng, Z. Zhu, and E. Gull, Solutions of the Two-Dimensional Hubbard Model: Benchmarks and Results from a Wide Range of Numerical Algorithms, Phys. Rev. X {\bf 5}, 041041 (2015).

\bibitem{qin20} M. Qin, C.-M. Chung, H. Shi, E. Vitali, C. Hubig, U. Schollwöck, S. R. White, and S. Zhang (Simons Collaboration on the Many-Electron Problem), Absence of Superconductivity in the Pure Two-Dimensional Hubbard Model, Phys. Rev. X {\bf 10}, 031016 (2020).

\bibitem{schafer21} T. Sch\"afer {\em et al.}, Tracking the Footprints of Spin Fluctuations: A MultiMethod, MultiMessenger Study of the Two-Dimensional Hubbard Model, Phys. Rev. X {\bf 11}, 011058  (2021).




\bibitem{lieb68} E. H. Lieb and F. Y. Wu, Absence of Mott Transition in an Exact Solution of the Short-Range, One-Band Model in One Dimension, Phys. Rev. Lett. {\bf 20}, 1445 (1968).





\bibitem{andersen84} O. K. Andersen and O. Jepsen, Explicit, First-Principles Tight-Binding Theory, Phys. Rev. Lett. {\bf 53}, 2571 (1984). 

\bibitem{sutton88} A.P. Sutton, M.W. Finnis, D.G. Pettifor, and Y. Ohta, The tight-binding bond model, J. Phys. C {\bf 21}, 35 (1988).


\bibitem{elstner98} M. Elstner, D. Porezag, G. Jungnickel, J. Elsner, M. Haugk, Th. Frauenheim, S. Suhai, and G. Seifert, Self-consistent-charge density-functional tight-binding method for simulations of complex materials properties, Phys. Rev. B {\bf 58}, 7260 (1998).

\bibitem{horsfield00} A. P. Horsfield and A. M. Bratkovsky, {\em Ab initio} tight binding, J. Phys.: Condens. Matter {\bf 12}, R1 (2000).

\bibitem{wang08} C.-Z. Wang, W.-C. Lu, Y.-X. Yao, J. Li, S. Yip, and K.-M. Ho, Tight-binding Hamiltonian from first-principles calculations, Sci. Model Simul. {\bf 15}, 81 (2008).

\bibitem{khan89} F. S. Khan and J. Q. Broughton, Simulation of silicon clusters and surfaces via tight-binding molecular dynamics, Phys. Rev. B {\bf 39}, 3688 (1989).

\bibitem{wang89} C. Z. Wang, C. T. Chan, and K. M. Ho, Empirical tight-binding force model for molecular-dynamics simulation of Si, Phys. Rev. B {\bf 39}, 8586 (1989).

\bibitem{goedecker94} S. Goedecker and L. Colombo, Efficient Linear Scaling Algorithm for Tight-Binding Molecular Dynamics, Phys. Rev. Lett. {\bf 73}, 122 (1994).

\bibitem{goodwin89} L. Goodwin, A. J. Skinner, and D. G. Pettifor, Generating Transferable Tight-Binding Parameters: Application to Silicon, Europhys. Lett. {\bf 9}, 701 (1989).



\bibitem{wannier37} G. H. Wannier, The Structure of Electronic Excitation Levels in Insulating Crystals, Phys. Rev. {\bf 52}, 191 (1937).

\bibitem{ashcroft76} N. W. Ashcroft and N. D. Mermin, {\em Solid State Physics} (Harcourt College Publishers, New York, 1976).

\bibitem{mazari97} N. Marzari and D. Vanderbilt, Maximally localized generalized Wannier functions for composite energy bands, Phys. Rev. B {\bf 56}, 12847 (1997).

\bibitem{marzari12} N. Marzari, A. A. Mostofi, J. R. Yates, I. Souza, and D. Vanderbilt, Maximally localized Wannier functions: Theory and applications, Rev. Mod. Phys. {\bf 84}, 1419 (2012).


\bibitem{kohn73} W. Kohn and J. R. Onffroy, Wannier Functions in a Simple Nonperiodic System, Phys. Rev. B {\bf 8}, 2485 (1973).

\bibitem{kivelson82} S. Kivelson, Wannier functions in one-dimensional disordered systems: Application to fractionally charged solitons, Phys. Rev. B {\bf 26}, 4269 (1982).

\bibitem{zhou10} S. Q. Zhou and D. M. Ceperley, Construction of localized wave functions for a disordered optical lattice and analysis of the resulting Hubbard model parameters, Phys. Rev. A {\bf 81}, 013402 (2010).

\bibitem{silvestrelli98} P. L. Silvestrelli, N. Marzari, D. Vanderbilt, M. Parrinello, Maximally-localized Wannier functions for disordered systems: Application to amorphous silicon, Solid State Commun. {\bf 107}, 7 (1998).

%

\bibitem{vidal91} F. J. Garca-Vidal, A. Martn-Rodero, F. Flores, J. Ortega, and R. P\'erez, Molecular-orbital theory for chemisorption: The case of H on normal metals, Phys. Rev. B {\bf 44}, 11412 (1991).

\bibitem{ortega98} J. Ortega, First-principles methods for tight-binding molecular dynamics,  Comput. Mater. Sci. {\bf 12}, 192 (1998).

\bibitem{spalek92} J. Spalek and W. W\'ojcik, Microscopic model of the Mott-Hubbard localization, Phys. Rev. B {\bf 45}, 3799 (1992).

\bibitem{spalek07} J. Spalek, E. M G\"orlich, A. Rycerz and R. Zahorbe\'nski, The combined exact diagonalization -- {\em ab initio} approach and its application to correlated electronic states and Mott–Hubbard localization in nanoscopic systems, J. Phys.: Condens. Matter {\bf 19},  255212 (2007).


\bibitem{hehre69} W. J. Hehre, R. F. Stewart, and J. A. Pople, Self-Consistent Molecular-Orbital Methods. I. Use of Gaussian Expansions of Slater-Type Atomic Orbitals, J. Chem. Phys. {\bf 51}, 2657 (1969).

%

\bibitem{chen-x} C. Chen and G.-W. Chern, unpublished. 

%

\bibitem{varshni57} Y. P. Varshni, Comparative Study of Potential Energy Functions for Diatomic Molecules, Rev. Mod. Phys. {\bf 29}, 664 (1957).

%

\bibitem{zeldovich44} Ya. B. Zeldovich and L. D. Landau, On the Relation between the Liquid and the Gaseous States of Metals, Zh. Eksp. Teor. Fiz. {\bf 14}, 32 (1944). 



\bibitem{ashcroft00} N. W. Ashcroft, { The hydrogen liquids,} Journal of Physics: Condensed Matter {\bf 12}, A129 (2000).

\bibitem{mcmahon12} J. M. McMahon, M. A. Morales, C. Pierleoni, and D. M. Ceperley, { The properties of hydrogen and helium under extreme conditions,} Review of Modern Physics {\bf 84}, 1607 (2012).


\bibitem{ackland15} G. J. Ackland, { Bearing down on hydrogen}, Science {\bf 348}, 1429 (2015).


\bibitem{bonev04} S. A. Bonev, E. Schwegler, T. Ogitsu, and G. Galli, { A quantum fluid of metallic hydrogen suggested by first-principles calculations,} Nature {\bf 431}, 669 (2004).

\bibitem{richardson97} C. F. Richardson and N. W. Ashcroft, { High temperatrue superconductivity in metallic hydrogen: electron-electron enhancements,} Physical Review Letters {\bf 78}, 118 (1997).

\bibitem{babaev04} E. Babaev, A. Sudbo,  and N. W. Ashcroft, { A superconductor to superfluid phase transition in liquid metallic hydrogen,} Nature {\bf 431}, 666 (2004).

\bibitem{cudazzo08} P. Cudazzo, G. Profeta, A. Sanna, A. Floris, A. Continenza, S. Massidda, and E. K. U. Gross, { Ab Initio description of high-temperature superconductivity in dense molecular hydrogen,} Physical Review Letters {\bf 100}, 257001 (2008).


\bibitem{wigner35} E. Wigner and H. B. Huntington, { On the possibility of a metallic modification of hydrogen,} Journal of Chemical Physics {\bf 3}, 764 (1935).

\bibitem{weir96} S. T. Weir, A. C. Mitchell, and W. J. Nellis, { Metallization of fluid molecular hydrogen at 140 GPa (1.4 Mbar),} Physical Review Letters {\bf 76}, 1860 (1996).

\bibitem{eremets11} M. I. Eremets and I. A. Troyan, { Conductive dense hydrogen,} Nature Materials {\bf 10}, 927 (2011).

\bibitem{dzyahura13} V. Dzyabura, M. Zaghoo, and I. F. Silvera, { Evidence of a liquid-liquid phase transition in hot dense hydrogen,} Proceedings of the National Academy of Sciences of USA {\bf 110}, 8040 (2013).

\bibitem{dias17} R. P. Dias and I. F. Silvera, { Observation of the Wigner-Huntington transition to metallic hydrogen,} Science {\bf 355} 715 (2017).

\bibitem{scandolo03} S. Scandolo, { Liquid-liquid phase transition in compressed hydrogen from first-principles simulations,} Proceedings of the National Academy of Sciences of USA {\bf 100}, 3051 (2003).

\bibitem{mazzola13} G. Mazzola, S. Yunoki, and S. Sorella, { Unexpectedly high pressure for molecular dissociation in liquid hydrogen by electronic simulation}, Nature Communications {\bf 5}, 3487 (2013).

\bibitem{morales13} M. A. Morales, J. M. McMahon, C. Pierleoni, and D. M. Ceperley, { Nuclear Quantum Effects and Nonlocal Exchange-Correlation Functionals Applied to Liquid Hydrogen at High Pressure,} Physical Review Letters {\bf 110}, 065702 (2013).

\bibitem{mcminis15} J. McMinis, R. C. Clay III, D. Lee, and M. A. Morales, { Molecular to atomic phase transition in hydrogen under high pressure,} Physical Review Letters {\bf 114}, 105305 (2015).

\bibitem{pierleoni16}, C. Pierleoni, M. A. Morales, G. Rillo, M. Holzmann, and D. M. Ceperley, { Liquid-liquid phase transition in hydrogen by coupled electron-ion Monte Carlo simulations}, Proceedings of the National Academy of Sciences of USA {\bf 113}, 4953 (2016).

\bibitem{mazzola18} G. Mazzola, R. Helled, and S. Sorella, { Phase Diagram of Hydrogen and a Hydrogen-Helium Mixture at Planetary Conditions by Quantum Monte Carlo Simulations,} Physical Review Letters {\bf 120}, 025701 (2018).




\bibitem{rose81} J. H. Rose, Metal-insulator transition in dilute alkali-metal systems, Phys. Rev. B {\bf 23}, 552 (1981).

\bibitem{chapman88} R. G. Chapman and N. H. March, Magnetic susceptibility of expanded fluid alkali metals, Phys. Rev. B {\bf 38}, 792 (1988).

\bibitem{redmer93} R. Redmer and W. W. Warren Jr. Magnetic susceptibility of Cs and Rb from the vapor to the liquid phase, Phys. Rev. B {\bf 48}, 14892 (1993).


%

\bibitem{warren89} W. W. Warren, Jr., G. F. Brennert, and U. El-Hanany, NMR investigation of the electronic structure of expanded liquid cesium, Phys. Rev. B {\bf 39}, 4038 (1989).

\bibitem{knuth97} B. Knuth, F. Hensel, and W. W. Warren Jr, Optical reflectivity and electron mass enhancement in expanded liquid caesium, J. Phys.: Condens. Matter {\bf 9}, 2693 (1997).

\bibitem{warren93} W. W. Warren Jr, Electron correlation and magnetism in dilute metals, J. Phys.: Condens. Matter {\bf 5}, B211 (1993).

%

\bibitem{falagan99} M. Reinaldo-Falag\'an, P. Tarazona, E. Chaco\'n, and J. P. Hernandez, Lattice-gas model driven by Hubbard electrons, Phys. Rev. E {\bf 60}, 2626 (1999).

\bibitem{falagan99b} M. Reinaldo-Falag\'an, J. P. Hernandez, E. Chaco\'n, and P. Tarazona, Gutzwiller approximation for a Hubbard lattice gas, J. Non-Cryst. Solids {\bf 250}, 20 (1999).

\bibitem{falagan03} M. Reinaldo-Falag\'an, P. Tarazona, E. Chaco\'n, E. Velasco, and J. P. Hernandez, Hard-sphere fluid with tight-binding electronic interactions: A glue model treatment, Phys. Rev. B {\bf 67}, 024209 (2003).

\bibitem{ishida02} I. Ishida, Effects of electron correlation on thermodynamic properties of expanded alkali fluids, J. Phys.: Condens. Matter {\bf 14}, 287 (2002).



\bibitem{tully12} J. C. Tully, Perspective: Nonadiabatic dynamics theory, J. Chem. Phys. {\bf 137}, 22A301 (2012).


\bibitem{hellmann37} H. Hellmann, {\em Einfhrung in die Quantenchemie}, (Deuticke, Leipzig, 1937).

\bibitem{feynman39} R. P. Feynman, Forces in Molecules, Phys. Rev. {\bf 56}, 340 (1939).








\bibitem{dahlen07} N. E. Dahlen and R. van Leeuwen, Solving the Kadanoff-Baym equations for inhomogeneous systems: Application to atoms and molecules, Phys. Rev. Lett. {\bf 98}, 153004 (2007).

\bibitem{friesen09} M. P. von Friesen, C. Verdozzi, and C.-O. Almbladh, Successes and Failures of Kadanoff-Baym Dynamics in Hubbard Nanoclusters, Phys. Rev. Lett. {\bf 103}, 176404 (2009).


\bibitem{schlunzen16} N. Schl\"unzen, S. Hermanns, M. Bonitz, and C. Verdozzi, Dynamics of strongly correlated fermions: {\em Ab initio} results for two and three dimensions, Phys. Rev. B {\bf 93}, 035107 (2016).

\bibitem{schlunzen20} N. Schl\"unzen, J.-P. Joost, and M. Bonitz, Achieving the Scaling Limit for Nonequilibrium Green Functions Simulations, Phys. Rev. Lett. {\bf 124}, 076601(2020).

%

\bibitem{bode11} N. Bode, S. V. Kusminskiy, R. Egger, and F. von Oppen, Scattering theory of current-induced forces in mesoscopic systems, Phys. Rev. Lett. {\bf 107}, 036804 (2011).

\bibitem{kershaw17} V. F. Kershaw and D. S. Kosov, Nonequilibrium Green's function theory for nonadiabatic effects in quantum electron transport, J. Chem. Phys. {\bf 147}, 224109 (2017).

\bibitem{dou17} W. Dou, G. Miao, and J. E. Subotnik, Born-Oppenheimer dynamics, electronic friction, and the inclusion of electron-electron interactions, Phys. Rev. Lett. {\bf 119}, 046001 (2017).

\bibitem{honeychurch19} T. D. Honeychurch and D. S. Kosov, Timescale separation solution of the Kadanoff-Baym equations for quantum transport in time-dependent fields, Phys. Rev. B {\bf 100}, 245423 (2019).

\bibitem{kershaw19} V. F. Kershaw and D. S. Kosov, Non-equlibrium Green's function theory for non-adiabatic effects in quantum transport: Inclusion of electron-electron interactions, J. Chem. Phys. {\bf 150}, 074101 (2019).




\bibitem{gull11} E. Gull, A. J. Millis, A. I. Lichtenstein, A. N. Rubtsov, M. Troyer, and P. Werner, Continuous-time Monte Carlo methods for quantum impurity models, Rev. Mod. Phys. {\bf 83}, 349  (2011).

\bibitem{bula08} R. Bulla, T. A. Costi, and T. Pruschke, Numerical renormalization group method for quantum impurity systems, Rev. Mod. Phys. {\bf 80}, 395 (2008).



\bibitem{georges92} A. Georges and G. Kotliar, Hubbard model in infinite dimensions, Phys. Rev. B {\bf 45}, 6479 (1992).

\bibitem{kajueter96} H. Kajueter and G. Kotliar, New Iterative Perturbation Scheme for Lattice Models with Arbitrary Filling, Phys. Rev. Lett. {\bf 77}, 131 (1996).




\bibitem{tang11} J. Tang, Y. Saad, A probing method for computing the diagonal of a matrix inverse, Numer. Linear Algebra Appl. {\bf 19}, 485 (2011).

\bibitem{carrier11} P. Carrier, J. M. Tang, Y. Saad, J. K. Freericks, Lanczos-based Low-Rank Correction Method for Solving the Dyson Equation in Inhomogenous Dynamical Mean-Field Theory, Phys. Procedia {\bf 15}, 22 (2011).



\bibitem{vilk97} Y. M. Vilk and A.-M. S. Tremblay, Non-Perturbative Many-Body Approach to the Hubbard Model and Single-Particle Pseudogap, J. Phys. I (France) {\bf 7}, 1309 (1997).


\bibitem{cheng22} Chen Cheng and Gia-Wei Chern, Enhancement of Atomic Diffusion due to Electron Delocalization in Fluid Metals, arXiv:2205.03888 (2022).

\end{thebibliography}
\end{document}